\documentclass[journal]{IEEEtran}
\usepackage{cite}
\usepackage{amsmath,amssymb,amsfonts}
\usepackage{graphicx}
\usepackage{textcomp}
\usepackage{balance}
\usepackage{xcolor}
\def\BibTeX{{\rm B\kern-.05em{\sc i\kern-.025em b}\kern-.08em
		T\kern-.1667em\lower.7ex\hbox{E}\kern-.125emX}}
\usepackage[ruled,vlined,linesnumbered]{algorithm2e}
\usepackage{mathtools}
\usepackage[T1]{fontenc}
\usepackage{url}
\usepackage{dsfont}
\usepackage{color}
\usepackage{bm}
\usepackage{mathrsfs}
\usepackage{multirow}
\usepackage{tudacolors}
\usepackage{hyperref}
\hypersetup{hidelinks}
\usepackage{diagbox}
\usepackage{framed}
\usepackage{stfloats}	% two-column floats cannot be placed at the bottom without this package
\usepackage{enumerate}
\usepackage{enumitem}
\usepackage{orcidlink}

\usepackage{subfigure}	% outdated package
\usepackage{pgfplotstable}
\usepackage{pgfplots}
\pgfplotsset{
	compat=newest,
	every tick label/.append style={font=\tiny},
	every axis plot/.append style={line width=0.5pt, mark size=2pt},	
	width=0.65\columnwidth,
	height=3.5cm,
	label style={font=\scriptsize},
}
\usetikzlibrary{plotmarks}
\usetikzlibrary{arrows.meta}
\usepgfplotslibrary{patchplots}
\usepackage{grffile}
\definecolor{MATLABcolor1}{rgb}{0.00000,0.44700,0.74100}%
\definecolor{MATLABcolor2}{rgb}{0.85000,0.32500,0.09800}%
\definecolor{MATLABcolor3}{rgb}{0.92900,0.69400,0.12500}%
\definecolor{MATLABcolor4}{rgb}{0.49400,0.18400,0.55600}%
\definecolor{MATLABcolor5}{rgb}{0.46600,0.67400,0.18800}%
\definecolor{MATLABcolor6}{rgb}{0.30100,0.74500,0.93300}%
\definecolor{MATLABcolor7}{rgb}{0.63500,0.07800,0.18400}%

\pgfplotscreateplotcyclelist{MATLABcolorlist}{
	MATLABcolor1, MATLABcolor2, MATLABcolor3, MATLABcolor4, MATLABcolor5, MATLABcolor6, MATLABcolor7
}

\pgfplotscreateplotcyclelist{MATLABcolormarklist}{
	{MATLABcolor1, mark=o, every mark/.append style = {solid}}, 
	{MATLABcolor2, mark=diamond, every mark/.append style = {solid}}, 
	{MATLABcolor3, mark=triangle, every mark/.append style = {solid,rotate=180}}, 
	{MATLABcolor4, mark=star, every mark/.append style = {solid}}, 
	{MATLABcolor5, mark=square, every mark/.append style = {solid}},
	{MATLABcolor6, mark=x, every mark/.append style = {solid}},
	{MATLABcolor7, mark=triangle, every mark/.append style = {solid}}
}

\usetikzlibrary{math}
\usetikzlibrary{calc}
\usetikzlibrary{shapes.misc}
\newcommand{\vect}[1]{\ensuremath{{\bm{#1}}}}
\newcommand{\mat}[1]{\ensuremath{{\bm{#1}}}}

\newcommand\bu{\ensuremath{\bm{u}}}

\newcommand\bx{\ensuremath{\bm{x}}}

\newcommand\bA{\ensuremath{\bm{A}}}

\newcommand\bD{\ensuremath{\bm{D}}}

\newcommand\bI{\ensuremath{\bm{I}}}

\newcommand\bR{\ensuremath{\bm{R}}}

\newcommand\bT{\ensuremath{\bm{T}}}

\newcommand\bX{\ensuremath{\bm{X}}}
\newcommand\bY{\ensuremath{\bm{Y}}}

\newcommand\bmu{\ensuremath{\bm{\mu}}}
\newcommand\bnu{\ensuremath{\bm{\nu}}}

\newcommand{\Real}{\ensuremath{\mathbb{R}}}
\newcommand{\Compl}{\ensuremath{\mathbb{C}}}
\newcommand{\Sym}{\ensuremath{\mathbb{S}}}

\newcommand{\Expect}{\ensuremath{\mathbb{E}}}
\renewcommand{\Pr}{\ensuremath{\mathbb{P}}}

\DeclareMathOperator{\tr}{tr}

\newcommand\imagunit{\ensuremath{\mathrm{j}}}
\newcommand\euler{\ensuremath{\mathrm{e}}}

\DeclarePairedDelimiter{\norm}{\lVert}{\rVert}
\DeclarePairedDelimiter{\abs}{\lvert}{\rvert}

\newcommand{\Trans}{\ensuremath{{\mathsf{T}}}}
\newcommand{\Herm}{\ensuremath{{\mathsf{H}}}}
\newcommand{\Frob}{\ensuremath{\mathsf{F}}}

\newcommand{\st}{\text{s.t.}}

\renewcommand{\vec}{\ensuremath{\operatorname{vec}}}
\newcommand\copyrighttext{%
	\footnotesize \textcopyright 2026 IEEE. 
	Personal use of this material is permitted. Permission from IEEE must be obtained for all other uses, in any current or future media, including reprinting/republishing this material for advertising or promotional purposes, creating new collective works, for resale or redistribution to servers or lists, or reuse of any copyrighted component of this work in other works. 
	DOI: \href{https://doi.org/10.1109/TSP.2026.3697636}{10.1109/TSP.2026.3697636}
}
\makeatletter
\def\ps@IEEEtitlepagestyle{
	\def\@oddfoot{\mycopyrightnotice}
	\def\@evenfoot{}
}
\def\mycopyrightnotice{
	{\footnotesize
		\begin{minipage}{\textwidth-2\fboxsep}%
			\centering%
			\noindent\fbox{\parbox{\linewidth}{\copyrighttext}}
		\end{minipage}
	}
}

\begin{document}
\title{Maximum A Posteriori Direction-of-Arrival Estimation via Mixed-Integer Semidefinite Programming}
\author{Tianyi Liu$^{\orcidlink{0000-0001-8338-1651}}$, Frederic Matter$^{\orcidlink{0000-0002-0499-1820}}$, Alexander Sorg, Marc E. Pfetsch$^{\orcidlink{0000-0002-0947-7193}}$, Martin Haardt$^{\orcidlink{0000-0001-7810-975X}}$,~\IEEEmembership{Fellow,~IEEE,} and Marius Pesavento$^{\orcidlink{0000-0003-3395-2588}}$,~\IEEEmembership{Senior Member,~IEEE}
	%	$^1$Communication Systems Group, Technische Universit\"at Darmstadt, Germany\\
	%	$^2$Research Group Optimization, Technische Universit\"at Darmstadt, Germany\\
	%	$^3$Communications Research Laboratory, Ilmenau University of Technology, Germany
    \thanks{This work was supported in part by the DFG PRIDE Project PE 2080/2-1, by the EXPRESS II Project within the DFG priority program CoSIP (DFG-SPP 1798), and by the Federal Ministry of Research, Technology and Space of Germany through the Open6GHub+ project (grant no. 16KIS2407).}
	\thanks{Tianyi Liu, Alexander Sorg, and Marius Pesavento are with the Communication Systems Group, Technical University of Darmstadt, 64283 Darmstadt, Germany (e-mail: tliu@nt.tu-darmstadt.de; alex@mail.ramanujan.eu; pesavento@nt.tu-darmstadt.de).
		%		This work was supported by the EXPRESS project within the DFG priority program CoSIP (DFG-SPP 1798) % project number: 273262315
		%		and the project ``Open6GHub'' (grant no. 16KISK014) sponsored by the Federal Ministry of Education and Research of Germany.
		%		%The authors acknowledge the financial support by the Federal Ministry of Education and Research of Germany in the project ``Open6GHub'' (grant no. 16KISK014).
	}
	\thanks{Frederic Matter and Marc E. Pfetsch are with the Research Group Optimization, Technical University of Darmstadt, 64283 Darmstadt, Germany (e-mail: frederic.matter@gmx.de; pfetsch@mathematik.tu-darmstadt.de).}
	\thanks{Martin Haardt is with the Communications Research Laboratory, Ilmenau University of Technology, 98684 Ilmenau, Germany (e-mail: martin.haardt@tu-ilmenau.de).}
	\thanks{Part of this work was presented at IEEE CAMSAP 2023~\cite{liuJointSparseEstimation2023}.}
}
\maketitle

\begin{abstract}
	%	The multiple measurement vectors (MMV) problem refers to the joint estimation of a row-sparse signal matrix from multiple realizations of mixtures with a known dictionary.
	%	%	multiple signal realizations where the signal samples share a common sparse support over a known dictionary.
	%	As a generalization of the standard sparse representation problem for a single measurement, this problem is fundamental in various applications in signal processing, e.g., spectral analysis and direction-of-arrival (DOA) estimation.
	% In this paper, we consider the maximum a posteriori (MAP) estimation for the multiple measurement vectors (MMV) problem with application to direction-of-arrival (DOA) estimation, which is classically formulated as a regularized least-squares (LS) problem with an $\ell_{2,0}$-norm constraint, and derive an equivalent mixed-integer semidefinite program (MISDP) reformulation.
	We propose a joint sparse maximum a posteriori (MAP) estimator for DOA estimation from multiple snapshots, reformulated as a mixed-integer semidefinite program (MISDP).
	% The proposed MISDP reformulation can be exactly solved by a generic MISDP solver using a semidefinite programming (SDP) based branch-and-bound method
	This enables efficient computation of globally optimal solutions using off-the-shelf MISDP solvers based on the branch-and-bound method.
	Unlike other nonconvex approaches for joint sparse recovery, such as the greedy methods and sparse Bayesian learning techniques, it provides a solution with an optimality assessment even with early termination.
	% We also present an approximate solution approach based on randomized rounding that yields high-quality feasible solutions of the proposed MISDP reformulation at a practically affordable computation time for large-scale problems.
	Additionally, we present a more scalable approximate solution approach for the MISDP problem based on randomized rounding.
	%	however, becomes computationally demanding for problems of extremely large dimensions. To further reduce the computation time in such scenarios, a relaxation-based approach can be employed to obtain an approximate solution of the MISDP reformulation, at the expense of a reduced estimation performance.
	Numerical simulations demonstrate the improved threshold behavior, resolution, and robustness of our proposed method against popular DOA estimation methods.
	In particular, the proposed method applied with the randomized rounding algorithm exhibits a superior estimation performance at a significantly reduced running time, compared to the deterministic maximum likelihood (DML) estimator.
	%	Moreover, unlike other nonconvex approaches for the MMV problem, including the greedy methods and the sparse Bayesian learning, the proposed MISDP-based method offers a guarantee of finding a global optimum.
\end{abstract}
\begin{IEEEkeywords}
	DOA estimation, multiple measurement vectors, joint sparsity, cardinality constraint, $\ell_{2,0}$-mixed-norm constraint, mixed-integer semidefinite program, maximum a posteriori estimation, randomized rounding
\end{IEEEkeywords}

\section{Introduction}
\label{sec:intro}

\IEEEPARstart{T}{he} multiple measurement vectors (MMV) problem is a fundamental challenge in signal processing and compressed sensing. It involves the joint estimation of multiple signals sharing a common sparse support over a known dictionary. The MMV problem arises in various applications, e.g., imaging~\cite{gorodnitskyNeuromagneticSourceImaging1995}, communications~\cite{fevrierReducedComplexityDecision1999,cotterSparseSolutionsLinear2005}, and signal processing~\cite{stoicaSpectralAnalysisSignals2005}.
The MMV problem is also known by several other names in the literature, e.g., simultaneous sparse coding~\cite{dongImageRestorationSimultaneous2015}, joint sparse coding~\cite{zhengAutomaticAnnotationSatellite2013}, and simultaneous sparse approximation~\cite{troppAlgorithmsSimultaneousSparse2006a,troppAlgorithmsSimultaneousSparse2006}.

%{\red Review on studies on MMV problem in the literature:}
Similar to the classical sparse signal reconstruction from a single measurement vector (SMV), the MMV problem is NP-hard due to the combinatorial nature of the cardinality constraint~\cite{natarajanSparseApproximateSolutions1995,davisAdaptiveGreedyApproximations1997}. Hence, approximate procedures are conventionally applied.
Many existing approximate solution approaches for the SMV case have been extended to the MMV case.
Those approaches can be roughly divided into greedy methods~\cite{cotterSparseSolutionsLinear2005,troppAlgorithmsSimultaneousSparse2006a,needellCoSaMPIterativeSignal2009,daviesRankAwarenessJoint2012}, convex relaxation approaches based on minimization of diversity measures~\cite{troppAlgorithmsSimultaneousSparse2006,malioutovSparseSignalReconstruction2005,steffensCompactFormulationEll2018,kowalskiSparseRegressionUsing2009,cotterSparseSolutionsLinear2005}, and sparse Bayesian learning methods \cite{wipfEmpiricalBayesianStrategy2007,gerstoftMultisnapshotSparseBayesian2016}.

The diversity minimization methods achieve sparse solutions by introducing in the objective function a penalty, referred to as the diversity measure, that is computationally convenient and encourages joint sparsity.
In particular, as a natural extension of basis pursuit~\cite{chenAtomicDecompositionBasis2001} or LASSO~\cite{tibshiraniRegressionShrinkageSelection1996} for the SMV case, the $\ell_{2,1}$-mixed-norm penalty is investigated in~\cite{cotterSparseSolutionsLinear2005,malioutovSparseSignalReconstruction2005}. An equivalent compact reformulation of the $\ell_{2,1}$-mixed-norm minimization, named SPARROW, is proposed in~\cite{steffensCompactFormulationEll2018}, which can be solved at a significantly reduced running time.
%asymptotic bias
%As regularization-based methods, the diversity minimization methods can be equivalently interpreted as maximum a posteriori (MAP) estimation with different priors under the framework of Bayesian inference~\cite{gribonvalShouldPenalizedLeast2011}.
The diversity minimization methods belong to the category of regularization-based methods, which can be equivalently interpreted as maximum a posteriori (MAP) estimators with different priors under the framework of Bayesian inference~\cite{gribonvalShouldPenalizedLeast2011}.

Another class of methods established in the Bayesian framework is known as sparse Bayesian learning (SBL). The parameters of the prior distribution are assumed to be known and considered as tuning parameters in the diversity minimization methods. In contrast, the SBL framework directly estimates the prior parameters by a type-II maximum likelihood, i.e., by maximizing the marginal likelihood that has been integrated over the parameter space~\cite{wipfEmpiricalBayesianStrategy2007}.
Although this marginal likelihood is multimodal, various iterative algorithms have been employed to efficiently obtain its stationary points, including the EM algorithm and other fixed-point methods~\cite{wipfEmpiricalBayesianStrategy2007,gerstoftMultisnapshotSparseBayesian2016}.
Recovery guarantees of several aforementioned methods for the MMV problem are established in~\cite{jinSupportRecoverySparse2013,chenTheoreticalResultsSparse2006,laiNullSpaceProperty2011,daviesRankAwarenessJoint2012,rameshSamplemeasurementTradeoffSupport2021,khannaSupportRecoveryJointly2022}.

%{\red Relation to DOA estimation:}
MMV-based parameter estimation is a classical problem in various applications in array signal processing including direction-of-arrival (DOA) estimation~\cite{krimTwoDecadesArray1996,vantreesOptimumArrayProcessing2002,zoubirArrayStatisticalSignal2014,pesaventoThreeMoreDecades2023,liuTwentyfiveYearsSensor2023}.
As a prominent class of DOA estimation methods, the subspace-based methods have been developed by exploiting the eigenstructure of the spatial correlation matrix, including the MUltiple Signal Classification (MUSIC) and ESPRIT along with their variants~\cite{schmidtMultipleEmitterLocation1986,barabellImprovingResolutionPerformance1983,royESPRITestimationSignalParameters1989,haardtUnitaryESPRITHow1995}.
However, in the case of correlated source signals and/or low sample sizes, the classical subspace-based methods experience a dramatic performance degradation as the signal subspace becomes rank deficient.
A common alternative approach that is known to be robust to the signal correlation is the deterministic maximum likelihood (DML) estimation, which is formulated as a nonlinear least-squares (LS) problem. The DML estimation has remarkable error performance in both the threshold and asymptotic region by fully exploiting the data model. Nevertheless, due to the nonlinearity and multimodality, the DML estimation is computationally expensive and generally requires a multidimensional grid search to obtain the exact solution.

Inspired by the capacity of compressed sensing~\cite{eldarCompressedSensingTheory2012}, sparsity-based DOA estimation methods have been developed. The DOA estimation from multiple snapshots is modeled as an MMV problem by introducing a predefined dictionary that is obtained from sampling the complete field-of-view (FOV)~\cite{yangSparseMethodsDirectionofarrival2018,liuTwentyfiveYearsSensor2023}.
The sparsity-based approach often exhibits excellent estimation performance in demanding scenarios at an affordable running time by using the aforementioned efficient MMV methods, such as the mixed-norm minimization~\cite{malioutovSparseSignalReconstruction2005,steffensCompactFormulationEll2018} and sparse Bayesian learning~\cite{gerstoftMultisnapshotSparseBayesian2016,nannuruSparseBayesianLearning2019}.
Under the SBL framework, several noise models other than the commonly used Gaussian model have been investigated in~\cite{mecklenbraukerRobustSparseMestimation2024} for improved robustness to outliers.

The preceding sparsity-based methods are built on the on-grid model, where the true directions are assumed to lie on the sampling grid. This assumption is often violated in practice, leading to performance degradation.
Gridless local search methods that find a local optimum of some fitting criterion, e.g., DML, starting from a good initial estimate can be used to refine the DOAs recovered by on-grid methods. Examples of gridless refinement methods include gradient methods~\cite{parkAtomconstrainedGridlessDOA2024,gerstoftAtomconstrainedMaximumLikelihood2025} and the alternating projection algorithm~\cite{ziskindMaximumLikelihoodLocalization1988}.
Alternative approaches for addressing the grid mismatch include off-grid methods that jointly model and estimate the grid mismatch errors~\cite{yangOffgridDirectionArrival2013,walewskiOffgridParameterEstimation2017,huangOffgridDirectionofarrivalEstimation2022,zamaniIterativeDictionaryLearningbased2016,tanIterativeAdaptiveDictionary2018} and gridless methods that operate in the continuous domain without discretization of the FOV~\cite{bhaskarAtomicNormDenoising2013,tangCompressedSensingGrid2013,yangGridlessSparseMethods2015,chiGuaranteedBlindSparse2016,chiHarnessingSparsityContinuum2020,steffensCompactFormulationEll2018,parkGridlessSparseCovariancebased2022,yangRobustStatisticallyEfficient2023,steffensGridlessCompressedSensing2017,liuGridlessParameterEstimation2024}.
The reader is referred to~\cite{yangSparseMethodsDirectionofarrival2018,liuTwentyfiveYearsSensor2023} for a review of sparsity-based DOA estimation methods.

\subsection{Our Contributions and Outline}
In this paper, we focus on the on-grid model that is applicable to arbitrary array geometries.
The relaxation of the cardinality measure in the mixed-norm minimization often leads to a degradation of the estimation quality and an asymptotic bias.
The SBL method provides more flexibility and avoids the overhead of tuning regularization parameters. However, its performance degrades dramatically in the cases with a large number of sources or high source correlations, as shown in the simulations in Section~\ref{sec:results}.

%Direction-of-Arrival (DOA) estimation aims at determining the direction of arrival of the incoming source signals using the measurements from an array of multiple sensors, which is a fundamental problem in array signal processing and plays a crucial role in various applications, such as radar, sonar, wireless communication, and microphone arrays~\cite{krimTwoDecadesArray1996,vantreesOptimumArrayProcessing2002,chungDOAEstimationMethods2014,pesaventoThreeMoreDecades2023}.

%defects of other methods in the literature: bias caused by the convex approximation $\ell_{2,1}$-norm

To overcome the above issues, we consider the MAP estimation with an uncorrelated Gaussian prior for joint sparse signal reconstruction from multiple measurement vectors, with application to DOA estimation.
%in array signal processing.
In contrast to the mixed-norm minimization approaches, we employ the exact $\ell_{2,0}$-norm constraint to avoid performance degradation caused by the relaxation of the cardinality constraint.
The corresponding MAP estimator is formulated as a regularized LS problem with an $\ell_{2,0}$-norm constraint,
which can be viewed as a generalization of the $\ell_0$-norm constrained LS problem investigated in~\cite{pilanciSparseLearningBoolean2015} from a single measurement to MMV case.
Compared to DML, the uncorrelated Gaussian prior introduces an additional Tikhonov regularization into the MAP estimation problem, which enables a reformulation to a mixed-integer semidefinite program (MISDP).
% can then be exactly reformulated as a mixed-integer semidefinite program (MISDP) by using the reformulation techniques in~\cite{pilanciSparseLearningBoolean2015}.
The presented MISDP reformulation can be viewed as an extension of the reformulation in~\cite{pilanciSparseLearningBoolean2015} to the MMV case. However, we provide a different interpretation that connects the reformulation to the concentration with respect to the nuisance parameters in the classical DOA estimation methods~\cite{pesaventoThreeMoreDecades2023}.

% The motivation for using the MAP estimator is that the MISDP problem class has recently been comprehensively studied in mathematical optimization.
The use of the MAP estimator is motivated by the recent comprehensive studies on the class of MISDP problems in mathematical optimization.
Many general-purpose solvers based on the state-of-the-art branch-and-bound method, such as SCIP-SDP~\cite{gallyFrameworkSolvingMixedinteger2018,matterPresolvingMixedIntegerSemidefinite2023,hojnyHandlingSymmetriesMixedInteger2023}, can be employed to obtain a global optimum of a MISDP problem more efficiently than the DML approach, which generally requires exhaustive search. Moreover, if the branch-and-bound solver is terminated early, an optimality assessment of the obtained solution is provided by the gap between the lower and upper bounds of the optimal value.
%The proposed MISDP reformulation can be solved by a generic MISDP solver such as SCIP-SDP~\cite{gallyFrameworkSolvingMixedinteger2018,matterPresolvingMixedIntegerSemidefinite2023,hojnyHandlingSymmetriesMixedInteger2023}, which, however, may become computationally expensive for problems of extremely large dimensions.
%To reduce the running time in such scenarios of large problem dimensions, we further develop an approximate solution approach for the proposed MISDP reformulation by generalizing the interval relaxation-based method proposed by Pilanci et al. in~\cite{pilanciSparseLearningBoolean2015} for the SMV case to the MMV case, which is, hence, referred to as Pilanci's method.
%Alternatively, to further reduce the computation time for problems of extremely large dimensions, one may employ the interval relaxation based approximate solution approach proposed by Pilanci et al. in~\cite{pilanciSparseLearningBoolean2015} for the MISDP reformulation in the SMV case.
% Due to the incorporation of a randomized rounding procedure in the branch-and-bound approach, the SCIP-SDP solver often quickly finds an optimal or nearly optimal solution, but the verification of the optimality of the obtained solution may still be computationally expensive for problems of extremely large dimensions.
% Motivated by the above observation,
To further improve scalability, we devise a randomized rounding algorithm as an extension of the solution approach of Pilanci et al. in~\cite{pilanciSparseLearningBoolean2015} for the single measurement case to the multiple measurement problem under consideration.
Randomized rounding provides a more efficient way to explore the integer feasible set and finds high-quality approximate solutions of the proposed MISDP reformulation at significantly reduced computation time for large-scale problems.

Simulation results demonstrate superior performance of our proposed methods in terms of threshold behavior, resolution, and robustness to heavy source correlations in comparison to several widely used DOA estimation methods.
In particular, compared to the DML estimator obtained by brute-force search over a multidimensional grid, the proposed MISDP-based method applied with the randomized rounding algorithm exhibits a superior error performance at a considerably reduced running time in difficult scenarios, e.g., in the cases with a limited number of snapshots, closely located sources, or highly correlated sources.
%Also, the proposed method provides an improved robustness to the source correlations and the increase of the number of sources when compared to subspace-based methods and the SBL method.
%On the other hand, it is observed that the interval relaxation based implementation fails in the case of a large sample size since the tightness of the interval relaxation is no longer satisfied.
%Nevertheless, in the case of very few snapshots, this relaxation-based algorithm can be used to find a satisfactory approximate solution of the MISDP reformulation at a greatly reduced running time.

In contrast to other nonconvex approaches, including the greedy methods and SBL, the proposed MISDP-based method offers a guarantee of finding a global optimum via a branch-and-bound solver.
In addition, using the MISDP reformulation, we extend the existing links between the considered $\ell_{2,0}$-norm constrained problem and its commonly used convex relaxation --- the $\ell_{2,1}$-norm minimization problem.

To summarize, the main contributions of this paper are:
\begin{itemize}
	\item \textit{Problem reformulation:}
		% We consider the MAP estimation of an MMV problem, which is formulated as a regularized LS problem with an $\ell_{2,0}$-norm constraint, and derive an equivalent MISDP reformulation.
		We propose a joint sparse MAP estimator for DOA estimation from multiple snapshots, which is reformulated as a MISDP problem.
		The presented MISDP reformulation can be viewed as an extension of the reformulation in~\cite{pilanciSparseLearningBoolean2015} to the MMV case. However, we provide a different interpretation that connects the reformulation to the concentration with respect to the nuisance parameters in the classical DOA estimation methods.
  The reformulation enables efficient computation of globally optimal solutions using state-of-the-art branch-and-bound solvers, with the added benefit of optimality assessment even if terminated early.
		% \item \textit{Efficient solution approach:} In contrast to the DML estimator, which is obtained by brute-force search, the MAP estimator, as the global optimum of the MISDP reformulation, can be achieved by an efficient MISDP solver using a semidefinite programming (SDP) based branch-and-bound method at a significantly reduced computation time. Our approach has the attractive feature that, if the branch-and-bound solver is terminated early, an optimality assessment of the obtained solution is provided by the gap between the lower and upper bounds of the optimal value.
		%	\item In the simulations, we also investigate the interval relaxation based low-cost approximate solution approach proposed by Pilanci et al. in~\cite{pilanciSparseLearningBoolean2015} for the MISDP reformulation in the SMV case. The simulation results reveal the failure of this approximate method in the case of a large sample size since the interval relaxation becomes less tight.
	\item \textit{Large problem handling:}
		% For problems of extremely large dimensions, we provide a randomized rounding algorithm that can find a satisfactory approximate solution of the proposed MISDP reformulation at a practically affordable computation time.
  To further improve scalability, we introduce a randomized rounding algorithm that finds high-quality approximate solutions of the MISDP problem at a practically affordable computation time for large-scale problems.
  Those approximate solutions are shown to provide superior estimation performance compared to the state-of-the-art methods in numerical simulations.
	\item \textit{Theoretical results:} Based on the MISDP reformulation, we extend the existing results on the theoretical links between the considered $\ell_{2,0}$-norm constrained formulation and the commonly used convex formulation with the $\ell_{2,1}$-norm regularization.
		%	\item We present a proper choice of the regularization parameter for the employed penalized least-squares by interpreting it as a maximum a posterior estimator with a Gaussian prior.
\end{itemize}

Lastly, we remark that, although, in this paper, the measurement model is described in an application of DOA estimation, the proposed method is not restricted to any specific dictionary structure and, therefore, can be directly applied to a joint sparse estimation problem in any other application.

The paper is organized as follows.
The sensor array signal model is presented in Section~\ref{sec:model}.
%In Section~\ref{sec:bayesian}, we present a proper choice of the regularization parameter by interpreting the penalized least-squares as a maximum a posteriori (MAP) estimator.
In Section~\ref{sec:bayesian}, we briefly review
%two conventional DOA estimation methods, namely
the DML estimator and the MAP estimator established in the Bayesian framework, as two classical multi-source estimation methods.
In Section~\ref{sec:problem}, the DOA estimation task is modeled as an MMV problem, and the equivalent MISDP reformulation is established.
Two solution approaches for the MISDP reformulation are then described in Section~\ref{sec:solution}.
In Section~\ref{sec:SPARROW}, using the developed reformulation, we provide a theoretical comparison between the $\ell_{2,0}$-norm constrained problem and the conventional convex method with $\ell_{2,1}$-norm minimization.
In Section~\ref{sec:generalization}, we briefly introduce a generalization of the proposed MISDP-based method for the DOA estimation with nonuniform source variances.
Simulation results are presented in Section~\ref{sec:results}, and conclusions are drawn in Section~\ref{sec:conclusion}.

\subsection{Notation}
We use $ x $, $ \bx $, and $ \bX $ to denote a scalar, column vector, and matrix, respectively.
%$\mathbb{R}$ and $ \Compl $ are the set of real and complex numbers.
%The letter~$\rm j$ is reserved to denote the imaginary unit.
%${\rm j}\!=\!\sqrt{-1}$.
%For any $x = \Compl$, $ \abs{x} $ denotes its magnitude, $ x^* $ its complex conjugate, and $ \Re (x) $ its real part.
For any $x \in \Compl$, $x^*$ denotes its complex conjugate.
%The Euclidean projection onto a set $ \set{X} $ is denoted by $ \opr{P}_{\set{X}}(\cdot) $.
%Those scalar operations are also used with matrix arguments, where they are applied in an elementwise manner.
The sets of $ M \times M $ Hermitian and positive semidefinite (PSD) Hermitian matrices are denoted by $ \Sym^M $ and $ \Sym_+^M $, respectively. The $M \times M$ identity matrix is denoted by $\vect{I}_M$, and $\vect{0}$ and $\vect{1}$ represent a zero matrix and all-ones matrix, respectively.
The symbols $ (\cdot)^\Trans$, $(\cdot)^\Herm$, and $(\cdot)^{-1} $ denote the transpose, Hermitian transpose, and inverse, respectively.
The trace operator is written as $\tr(\cdot)$.
The Frobenius norm and the $\ell_{p,q}$-mixed-norm of a matrix $\vect{X}$ are referred to as $ \norm{\vect{X}}_\Frob $ and $\norm{\vect{X}}_{p,q}$, respectively, while the $\ell_p$-norm of a vector is defined as $\norm{\vect{x}}_p$. In particular, the $\ell_0$-pseudo-norm $\norm{\vect{x}}_0$ counts the number of nonzero entries in the vector $\vect{x}$.
%For a vector $\vect{x}$, the $k$th entry is $x_k$.
%Also, the $k$th entry of a vector $\vect{x}_l$ that itself carries a subscript will be denoted by $x_{kl}$.
%The Hadamard and Kronecker products are denoted by $\odot$ and $\otimes$, respectively;
%	symbol $ \bzero $ is a zero matrix.

\section{Signal Model}
\label{sec:model}

\begin{figure}[t]
	\centering
	% Macro files
%\input{figures/texMacros}
\input{figures/tikzMacros}
%\definecolor{mycolor1}{RGB}{136, 0, 21}%
%\definecolor{mycolor2}{RGB}{185,122,187}%
%%\definecolor{mycolor3}{RGB}{237,28,36}%
%\definecolor{mycolor3}{RGB}{255,32,27}%
%\definecolor{mycolor4}{RGB}{255,174,201}%
%\definecolor{mycolor5}{RGB}{255,137,29}%
%\definecolor{mycolor6}{RGB}{255,201,14}%
%\definecolor{mycolor7}{RGB}{255,242,0}%
%\definecolor{mycolor8}{RGB}{239,228,176}%
%%\definecolor{mycolor9}{RGB}{34,177,76}%
%\definecolor{mycolor9}{RGB}{31,194,43}%
%\definecolor{mycolor10}{RGB}{181,230,29}%
%%\definecolor{mycolor11}{RGB}{0,162,232}%
%\definecolor{mycolor11}{RGB}{0,119,251}%
%%\definecolor{mycolor12}{RGB}{153,217,234}%
%\definecolor{mycolor12}{RGB}{94,243,255}%
%
%\definecolor{mycolor13}{RGB}{63,72,204}%
%\definecolor{mycolor14}{RGB}{112,146,190}%
%\definecolor{mycolor15}{RGB}{163,73,164}%
%\definecolor{mycolor16}{RGB}{200,191,231}%
\begin{tikzpicture}[scale=1]  
\draw [opacity = 0] (-3.,-0.75) rectangle (3,2.5);  
\scriptsize
\tikzset{>=stealth}

%%% draw grid
\draw[lightgray,thin] (2,0) arc (0:180:2);
\foreach \phi in {0,10,...,180} {
	\draw[lightgray,dashed,thin] (0,0) -- (\phi:2);
}
\draw[lightgray,dashed,thin] (0,0) -- (0,0);
%\node[anchor=east] at (-2,0) {$\nu_1$};
%\node[anchor=west] at (2,0) {$\nu_K$};
%\draw[->,lightgray,thin] (173:2.3) arc (173:155:2.3);
\draw[rounded corners = 2pt, fill = lightgray] (-2.25, -0.25) rectangle ++(4.5, 0.25);
% array 1
% draw nodes
\foreach \x in {0,1,2,5} {
	\node[sensor, fill=black] (s\x) at (0.8*\x-2,-0.125) {};
%	\node[below=s\x] {$\xi$};
}
	\node[sensor, fill=black] (s3) at (0.8*3-1.5,-0.125) {};
	\node[sensor, fill=black] (s4) at (0.8*4-2,-0.125) {};
	\node[below, yshift=-5pt] at (s0) {$\xi_1$};
	\node[below, yshift=-5pt] at (s1) {$\xi_2$};
	\node[below, yshift=-5pt] at (s2) {$\xi_3$};
	\node[below, yshift=-5pt] at (s5) {$\xi_M$};
	\node[below, yshift=-7pt] at (s3) {\large$\dots$};
%\node[anchor=north] at (0,-0.1cm) {Sensor Array};

% draw connections
\draw[gray] (s0) -- (s1) -- (s2) -- (s3) -- (s4) -- (s5); % A <-> C
%\node[anchor = north] at ($(s0)+(0, -0.1cm)$){Sensor $1$};
%\node[anchor = north] at ($(s5)+(0, -0.1cm)$){Sensor $M$};
\node[anchor=north] at (0,-0.75cm) {Arbitrary linear array with $M$ sensors};
%%% draw coordinate system
\node[circle, inner sep=2pt, anchor=center] (cs) at (0,0) {};

%%% draw sources
\def\angA{180-40}
\def\angB{180-70}
\def\angC{180-90}
\def\angD{180-130}
\def\angE{180-85}

\coordinate (src1) at (intersection cs: first line={(cs)--($(cs)+(\angA:6)$)}, second line={(0,2)--(3,2)}) {};
\coordinate (src2) at (intersection cs: first line={(cs)--($(cs)+(\angB:6)$)}, second line={(0,2)--(3,2)}) {};
\coordinate (src3) at (intersection cs: first line={(cs)--($(cs)+(\angC:6)$)}, second line={(0,2)--(3,2)}) {};
\coordinate (src4) at (intersection cs: first line={(cs)--($(cs)+(\angD:6)$)}, second line={(0,2)--(3,2)}) {};
\coordinate (src5) at (intersection cs: first line={(cs)--($(cs)+(\angE:6)$)}, second line={(0,1.9)--(3,1.9)}) {};

%TUDa-2b, TUDa-4b, TUDa-8b, TUDa-11b

\draw[TUDa-2b,->] (src1.center) -- (cs.center); % A <-> C
\draw[TUDa-4b,->] (src2.center) -- (cs.center); % A <-> C
\draw[TUDa-8b,->] (src3.center) -- (cs.center); % A <-> C
\draw[TUDa-11b,->] (src4.center) -- (cs.center); % A <-> C

\node[source,fill=TUDa-2b, label={Source $1$}] at (src1.center) {};
\node[source,fill=TUDa-4b] at (src2.center) {};
\node[source,fill=TUDa-8b] at (src3.center) {};
\node[source,fill=TUDa-11b, label={Source $L$}] at (src4.center) {};
\node[anchor = south, label={above:{\large$\dots$}}] at (src5.center) {};

\draw[gray,thin,->] ($(0.4,0)$) arc (0:140:0.4);
\node [black,anchor=north east] at (-0.1, 0.75) {$\theta_1$};
\draw[gray,thin,->] ($(0.5,0)$) arc (0:50:0.5);
\node [black,anchor=north east] at (1, 0.5) {$\theta_{L}$};
\end{tikzpicture}   
	\caption{Exemplary setup for a linear array of $M$ sensors and $L$ source signals.}
	\label{fig:array}
\end{figure}

As depicted in Fig.~\ref{fig:array}, we consider a linear array of $M$ omnidirectional sensors.
%Let $\zeta_m \in \Real$ for $m=1,\ldots,M$ be the position of the $m$th sensor in half signal wavelength.
Assume that $L$ narrowband far-field source signals impinge from distinct directions $\theta_1,\ldots,\theta_L \in [0,180^\circ]$. The corresponding spatial frequencies are
\begin{equation}
	\mu_l = \pi \cos (\theta_l) \in [-\pi, \pi)
\end{equation}
for $l=1,\ldots,L$ and summarized in the vector $\mat{\mu} = [\mu_1,\ldots,\mu_L]^\Trans$.
We consider the DOA estimation with multiple snapshots, where the array output provides measurements recorded at $N$ time instants. The sources emit time-varying signals, whereas the spatial frequencies in $\mat{\mu}$ remain constant. Let $\mat{Y} = [\vect{y}_1,\ldots,\vect{y}_N] \in \Compl^{M \times N}$ contain the $N$ snapshots and the $(m,n)$th entry $y_{m,n}$ is the output of sensor $m$ at time instant $n$.
The measurement matrix is modeled as
\begin{equation}\label{eq:model}
	\vect{Y} = \vect{A} (\bmu)   \vect{\Psi} + \vect{N}, % = \vect{X}_{0} + \vect{N}
\end{equation}
where $\vect{\Psi} = [\vect{\psi}_1,\ldots,\vect{\psi}_N] \in \Compl^{L \times N}$ is the source waveform with $ \psi_{l,n} $ being the signal emitted by source $ l $
%and impinging on the sensor array
at time instant $ n $.
The matrix $\vect{A} (\bmu)$ contains $L$ steering vectors as
\begin{equation}\label{eq:steerMat}
	\vect{A} (\bmu) =
	\begin{bmatrix}
		\vect{a}(\mu_{1}) & \ldots & \vect{a}(\mu_{L})
	\end{bmatrix} \in \mathbb{C}^{M\times L},
\end{equation}
where
%\begin{equation}\label{key}
$\mat{a}(\mu) =
[\euler^{\imagunit \mu \xi_1} , \ldots, \euler^{\imagunit \mu \xi_M}]^\Trans$
%\end{equation}
is the steering vector for the frequency $\mu$ and $\xi_1,\ldots,\xi_M$ denote the sensor locations in the linear array measured in half-wavelength.
The matrix $\vect{N} = [\vect{n}_1,\ldots,\vect{n}_N] \in \Compl^{M \times N}$ represents independent and identically distributed (i.i.d.) circular and spatio-temporal white Gaussian noise with variance $ \sigma^2 $ of each noise entry $n_{m,n}$.

\section{Deterministic Maximum Likelihood and Maximum A Posteriori Estimators}
\label{sec:bayesian}

%Apparently, the choice of the regularization parameter $\rho$ in~\eqref{prob:MMVL20_original} has a critical effect on the performance of the proposed method.
%The penalized least-squares regression is commonly interpreted in a Bayesian framework as a maximum a posteriori (MAP) estimator with the penalty function being the negative logarithm of the prior~\cite{gribonvalShouldPenalizedLeast2011}. Therefore, in the following, we present an appropriate choice of $\rho$ by establishing a connection between the formulation in~\eqref{prob:MMVL20IP1} and the MAP estimator with an uncorrelated Gaussian prior.

% In this section
%, based on the signal model in~\eqref{eq:model},
We briefly review
%two conventional DOA estimation methods, namely
the DML estimator and the MAP estimator established in the Bayesian framework, as two classical multi-source estimation methods.
As they are computationally demanding if the number of sources is large, we propose an equivalent reformulation of the MAP estimation in Section~\ref{sec:problem}. The resulting MISDP reformulation enables a computationally efficient solution to the MAP estimation using state-of-the-art numerical MISDP solvers.

In the DML approach, the source waveform matrix $\vect{\Psi}$ in~\eqref{eq:model} is considered deterministic and unknown. According to the signal model~\eqref{eq:model}, the snapshots $\vect{y}_n$ are statistically independent and complex normally distributed with mean $\vect{A}(\vect{\mu}) \vect{\psi}_n$ and covariance $\sigma^2 \vect{I}_M$, i.e.,
\begin{equation} \label{eq:likelihood}
	\vect{y}_n | \vect{\psi}_n \sim \mathcal{CN} (\vect{A}(\vect{\mu}) \vect{\psi}_n, \sigma^2 \vect{I}_M).
\end{equation}
Thus, the DML estimator for frequencies $\vect{\mu}$ and source waveforms $\vect{\Psi}$ is obtained from the nonlinear LS problem~\cite{vantreesOptimumArrayProcessing2002}
\begin{equation} \label{prob:DML}
	\min_{\vect{\mu} \in [-\pi, \pi)^L, \, \vect{\Psi} \in \Compl^{L \times N}} \quad \norm{\vect{A}(\vect{\mu}) \vect{\Psi} - \vect{Y}}_\Frob^2.
\end{equation}
%where $\|\cdot\|_\Frob$ denotes the Frobenius norm.
Since we are interested in estimating the DOA parameters $\vect{\mu}$, the objective function~\eqref{prob:DML} is concentrated with respect to the nuisance parameters $\vect{\Psi}$. That is, for each $\vect{\mu}$, the minimizer of the nuisance parameters $\vect{\Psi}$ is expressed in closed form, which is then substituted into the original objective function to obtain the concentrated optimization problem.
Particularly, the DML estimation problem~\eqref{prob:DML} is concentrated as~\cite{pesaventoThreeMoreDecades2023}:
\begin{equation} \label{prob:concentratedDML}
	\min_{\vect{\mu} \in [-\pi,\pi)^L} \quad \tr \left(\vect{Y}^\Herm \vect{\Pi}_{\vect{A}(\vect{\mu})}^\perp \vect{Y} \right),
\end{equation}
where $\vect{\Pi}_{\vect{A}(\vect{\mu})}^\perp = \vect{I}_M - \vect{A} (\vect{\mu}) \left(\vect{A}(\vect{\mu})^\Herm \vect{A}(\vect{\mu}) \right)^{-1} \vect{A}(\vect{\mu})^\Herm $ denotes the orthogonal projector onto the orthogonal complement of the column space of the matrix $\vect{A}(\vect{\mu})$.

%For simplicity, we first consider the case with a single measurement vector. Let $\vect{x}$ be a signal vector at some time instant.
The maximum a posteriori (MAP) estimator~\cite{zoubirArrayStatisticalSignal2014,gribonvalShouldPenalizedLeast2011} is another widely used estimation method that is closely related to the ML estimation.
In this approach, only the DOAs are considered deterministic, whereas the source waveforms are assumed to be stochastic.
%Unlike in the DML estimator, where $\vect{x}$ is assumed to be deterministic, here
We consider the spatio-temporal i.i.d. assumption that the signal waveforms $\psi_{l,n}$ are statistically independent for different sources and snapshots. They follow the same circularly-symmetric complex Gaussian distribution
\begin{equation}\label{eq:prior}
	\vect{\psi}_n \sim \mathcal{CN} (\vect{0},\gamma \vect{I}_L),
\end{equation}
where $\gamma$ is the source power that is assumed to be known a priori.
%That is, the signals from all possible directions are assumed to be uncorrelated and have the same average power, denoted by $P_\Psi$.
%When $L$ directions have been chosen from the dictionary $\vect{A}$, i.e., the binary vector $\vect{u}$ in~\eqref{prob:MMVL20IP1} is fixed,
%According to the signal model in~\eqref{eq:model}, the conditional distribution of each snapshot $\vect{y}_n$ given the corresponding source waveform $\vect{\psi}_n$ is
%\begin{equation}\label{key}
%	\vect{y}_n | \vect{\psi}_n \sim \mathcal{CN} (\vect{A}(\vect{\mu}) \vect{\psi}_n, \sigma^2 \vect{I}_M).
%\end{equation}
By the Bayes' rule, the MAP estimator for the uncorrelated Gaussian prior~\eqref{eq:prior} is given by the solution of the following regularized LS problem~\cite{gribonvalShouldPenalizedLeast2011}:
\begin{subequations} \label{prob:MAP}
	\begin{align}
		& \max_{\vect{\mu} \in [-\pi, \pi)^L, \, \vect{\Psi} \in \Compl^{L \times N}} \ \prod_{n=1}^N p (\vect{\psi}_n | \vect{y}_n) \nonumber \\
		\Leftrightarrow & \max_{\vect{\mu} \in [-\pi, \pi)^L, \, \vect{\Psi} \in \Compl^{L \times N}} \ \prod_{n=1}^N p(\vect{y}_n|\vect{\psi}_n) p(\vect{\psi}_n) \\
		\Leftrightarrow & \min_{\vect{\mu} \in [-\pi, \pi)^L, \, \vect{\Psi} \in \Compl^{L \times N}} \ \sum_{n=1}^N - \log (p (\vect{y}_n | \vect{\psi}_n)) - \log (p(\vect{\psi}_n)) \nonumber \\
		\Leftrightarrow & \min_{\vect{\mu} \in [-\pi, \pi)^L, \, \vect{\Psi} \in \Compl^{L \times N}} \ \norm{\vect{A} (\vect{\mu}) \vect{\Psi} - \vect{Y}}_\Frob^2 + \rho \norm{\vect{\Psi}}_\Frob^2, \label{prob:MAP_LS}
	\end{align}
\end{subequations}
where $p(\cdot)$ denotes the probability density function and
\begin{equation} \label{eq:rho}
	\rho = {\sigma^2}/{\gamma}.
\end{equation}
The first LS data fitting term~\eqref{prob:MAP_LS} resulting from the likelihood $p(\vect{y}_n | \vect{\psi}_n)$ is identical to the DML cost function~\eqref{prob:DML},
%In~\eqref{prob:MAP}, The distribution $p(\vect{y}_n | \vect{\psi}_n)$  is the likelihood as used for the DML estimation in~\eqref{prob:DML}, which similarly leads to a LS data fitting term,
whereas the prior distribution $p(\vect{\psi}_n)$~\eqref{eq:prior}, introduces the Tikhonov regularization term~\eqref{prob:MAP_LS}.
%Compared to the DML estimation, the additional Tikhonov regularization in the MAP estimation enables the MISDP reformulation that will be presented in Section~\ref{sec:problem}.
%Comparing~\eqref{prob:MMVL20IP1} and~\eqref{eq:MAP}, we can conclude that, when the regularization parameter $\rho$ is chosen to be
%\begin{equation}\label{eq:rho}
%	\rho = \frac{\sigma^2}{P_\Psi},
%\end{equation}
%problem~\eqref{prob:MMVL20IP1}, or equivalently problem~\eqref{prob:MMVL20_original}, corresponds to a MAP estimator with the Gaussian prior in~\eqref{eq:prior} and an additional group-sparsity assumption in the case with $N$ independent snapshots.
%Additionally, when the regularization in~\eqref{prob:MMVL20_original} is discarded, the problem becomes a discretized DML estimation, i.e., the DML estimation over a grid.
For a given $\vect{\mu}$, the minimizer of the nuisance parameters $\vect{\Psi}$ in~\eqref{prob:MAP} admits the well-known Tikhonov closed-form solution~\cite{boydConvexOptimization2004}
\begin{equation} \label{eq:partSol_MAP}
	\widetilde{\vect{\Psi}} = \left(\vect{A}(\vect{\mu})^\Herm \vect{A}(\vect{\mu}) + \rho \vect{I}_L \right)^{-1} \vect{A}(\vect{\mu})^\Herm \vect{Y}.
\end{equation}
Then, by substituting~\eqref{eq:partSol_MAP} into~\eqref{prob:MAP}, the MAP estimation is concentrated as
\begin{equation} \label{prob:concentratedMAP1}
	\min_{\vect{\mu} \in [-\pi, \pi)^L} \quad \tr \left(\vect{Y}^\Herm \widetilde{\vect{\Pi}}_{\vect{A}(\vect{\mu})}^\perp \vect{Y}\right)
\end{equation}
with $\widetilde{\vect{\Pi}}_{\vect{A}(\vect{\mu})}^\perp = \vect{I}_M - \vect{A} (\vect{\mu}) \left(\vect{A}(\vect{\mu})^\Herm \vect{A}(\vect{\mu}) + \rho \vect{I}_L \right)^{-1} \vect{A}(\vect{\mu})^\Herm $.
By using the matrix inversion lemma, the matrix $ \widetilde{\vect{\Pi}}_{\vect{A}(\vect{\mu})}^\perp $ is rewritten as $\widetilde{\vect{\Pi}}_{\vect{A}(\vect{\mu})}^\perp = (\tfrac{1}{\rho} \vect{A}(\vect{\mu}) \vect{A} (\vect{\mu})^\Herm + \vect{I}_M )^{-1}$, which leads to the following equivalent expression of the concentrated MAP estimation in~\eqref{prob:concentratedMAP1}:
\begin{equation} \label{prob:concentratedMAP2}
	\min_{\vect{\mu} \in [-\pi, \pi)^L} \quad \tr \left(\vect{Y}^\Herm (\tfrac{1}{\rho} \vect{A}(\vect{\mu}) \vect{A} (\vect{\mu})^\Herm + \vect{I}_M )^{-1} \vect{Y}\right).
\end{equation}

%\section{Time Compression Using Singular Value Decomposition}
%
%\begin{equation}
%	\vect{R} = \Expect \left[\vect{y}_n \vect{y}_n^\Herm\right] = \vect{A}(\vect{\mu}) \vect{P} \vect{A}(\vect{\mu})^\Herm + \sigma_\text{n}^2 \vect{I}
%\end{equation}
%\begin{equation}
%	\vect{R} = \vect{U} \vect{\Lambda} \vect{U}^\Herm = \vect{U}_\text{s} \vect{\Lambda}_\text{s} \vect{U}_\text{s}^\Herm + \vect{U}_\text{n} \vect{\Lambda}_\text{n} \vect{U}_\text{n}^\Herm
%\end{equation}
%\begin{equation}
%	\vect{Y} = \widehat{\vect{U}} \widehat{\vect{\Sigma}} \widehat{\vect{V}}^\Herm
%\end{equation}
%\begin{equation}
%	\vect{Y}_\text{s} = \widehat{\vect{U}} \widehat{\vect{\Sigma}} \vect{J}_\text{s} = \vect{Y} \widehat{\vect{V}} \vect{J}_\text{s}
%\end{equation}
%$\vect{J}_\text{s} = \in \Real^{N \times L}$
%\begin{equation}
%	\vect{Y}_\text{s} = \vect{A} (\vect{\mu}) \vect{\Psi}_\text{s} + \vect{N}_\text{s}
%\end{equation}

\section{A MISDP Reformulation of MAP Estimation for the MMV Problem}
\label{sec:problem}

Due to the quadratic term in the matrix inversion, both the DML~\eqref{prob:concentratedDML} and the MAP~\eqref{prob:concentratedMAP2} estimation problems are nonconvex and multimodal with a large number of local minima. Hence, the corresponding optimization procedure is computationally demanding and requires a multidimensional grid search to find the solution.
%Therefore, in this section, following the same line of analysis as in~\cite{yangSparseMethodsDirectionofarrival2018}, we model the above DOA estimation task as an MMV problem by introducing a predefined dictionary that samples the complete field-of-view (FOV).
Inspired by compressed sensing~\cite{eldarCompressedSensingTheory2012}, the above DOA estimation can be modeled as an MMV problem by introducing a predefined dictionary that samples the complete FOV~\cite{yangSparseMethodsDirectionofarrival2018}.
% In this section, we first introduce the MMV-based model for DOA estimation.
For the MMV problem, a dictionary-based MAP estimation problem is developed according to~\eqref{prob:MAP}, which is reformulated as a MISDP problem by the reformulation techniques in~\cite{pilanciSparseLearningBoolean2015,steffensCompactFormulationEll2018}.

Recovering the frequencies $\vect{\mu}$ from the measurement matrix $\vect{Y}$ is formulated as an MMV problem by exploiting the sparse representation for the model~\eqref{eq:model}:
\begin{equation} \label{eq:sparseModel}
	\vect{Y} = \vect{A} (\bnu ) \vect{X} + \vect{N},
\end{equation}
where $\vect{A} (\bnu ) = [\vect{a}(\nu_1),\ldots,\vect{a}(\nu_K)] \in \Compl^{M \times K}$ is an overcomplete dictionary constructed by sampling the FOV in $K \gg L$ directions with spatial frequencies $ \bnu = [\nu_1,\ldots,\nu_K]^\Trans $ and $ \vect{X} \in \Compl^{K \times N} $ is a sparse representation of the source signal matrix $ \vect{\Psi} $.
Specifically, provided that the true frequencies $ \bmu $ are contained in the frequency grid, i.e.,
\begin{equation} \label{eq:on-grid}
	\{ \mu_l\}_{l=1}^{L} \subset \{\nu_k\}_{k=1}^K,
\end{equation}
then $\vect{X} = [\vect{x}_1,\ldots,\vect{x}_K]^\Trans$ admits a row-sparse structure, which has $L$ nonzero rows corresponding to the signal waveforms of the $ L $ sources, i.e., $\vect{A}(\vect{\mu}) \vect{\Psi} = \vect{A}(\vect{\nu}) \vect{X}$.
%i.e., only rows corresponding to the original spatial frequencies are active.
Thus, the DOA estimation is an MMV problem, which aims at jointly recovering a set of signal samples in $ \vect{X} $ that have common sparse support over a given dictionary $ \vect{A}(\vect{\nu}) $ from multiple measurement vectors in~$ \vect{Y} $.
The frequencies are then estimated from the support of the recovered row-sparse matrix $\widehat{\vect{X}} = [\widehat{\vect{x}}_1,\ldots,\widehat{\vect{x}}_K]^\Trans$ by
%\begin{equation}
$\{\hat{\mu}_l\}_{l=1}^L = \{ \nu_k \mid \norm{\widehat{\vect{x}}_k}_0 > 0, \, k=1,\ldots,K \}$.
%\end{equation}
%	where the $\ell_0$-pseudo-norm $\norm{\widehat{\vect{x}}_k}_0$ counts the number of nonzeros entries in $\widehat{\vect{x}}_k$.
For simplicity, the dictionary is referred to as $ \vect{A} = \vect{A} (\bnu) $.

The $\ell_{p,q}$-mixed-norms enforce the row-sparsity in sparse reconstruction problems~\cite{kowalskiSparseRegressionUsing2009}.
The $\ell_{p,q}$-norm for a matrix $\vect{X} = [\vect{x}_1,\ldots,\vect{x}_K]^\Trans$ is defined as
\begin{equation*} \label{eq:pq-norm}
	\norm{\vect{X}}_{p,q} = \norm{\vect{x}^{(\ell_p)}}_q
	\quad \text{with} \quad
	\vect{x}^{(\ell_p)} =
	\begin{bmatrix}
		\norm{\vect{x}_1}_p \cdots \norm{\vect{x}_K}_p
	\end{bmatrix}^\Trans.
\end{equation*}
The inner $\ell_p$-norm applied to each row provides a nonlinear coupling among the elements in a row, whereas the outer $\ell_q$-norm applied to the norms of all rows approximately measures the row-sparsity.
The $\ell_{p,0}$-pseudo-norm represents the exact number of nonzero rows, which, however, typically leads to an NP-hard problem due to its nonconvexity. In~\cite{malioutovSparseSignalReconstruction2005}, the $\ell_{2,1}$-norm is utilized as a convex approximation of the $\ell_{2,0}$-norm, to address the MMV problem described above.
In contrast, we consider the exact MAP estimation for the sparse model~\eqref{eq:sparseModel}.

Let us impose the same spatio-temporal i.i.d. zero-mean complex Gaussian prior assumption~\eqref{eq:prior} on the nonzero rows of $\vect{X}$. The entries $x_{k,n}$ in the nonzero rows of $\vect{X}$ are independent across snapshots and DOAs, and follow the distribution
\begin{equation} \label{eq:prior_sparse}
	x_{k,n} \sim \mathcal{CN} (0, \gamma)
\end{equation}
for $n = 1,\ldots,N$.
Similar to~\eqref{prob:MAP}, the MAP estimator for the sparse model~\eqref{eq:sparseModel} is given by the following regularized LS problem with $\ell_{2,0}$-norm constraint:
\begin{align}
	\label{prob:MMVL20_original}
	\underset{\substack{\bX \in \Compl^{K \times N}, \, \norm{\bX}_{2,0} \leq L}}{\min} \quad  \norm{\bA \bX - \bY}_\Frob^2 +
	\rho \norm{\bX}_\Frob^2.
\end{align}
%where $\|\cdot\|_\Frob$ denotes the Frobenius norm.
The DML approach for the sparse model in~\eqref{eq:sparseModel} is obtained from~\eqref{prob:MMVL20_original} by choosing the regularization parameter $\rho= 0$.
The problem~\eqref{prob:MMVL20_original} is equivalent to solving the MAP estimation problem~\eqref{prob:MAP_LS} via a brute-force search over the spatial frequency grid~\eqref{eq:on-grid}.
% To be consistent with~\eqref{prob:MAP_LS}, $\vect{X}$ is not restricted to have exactly $L$ nonzero rows, i.e., $\norm{\vect{X}}_{2,0} = L$, since the waveforms in~\eqref{prob:MAP_LS} can also attain zero values for any chosen combination of spatial frequencies $\vect{\mu}$.
However, compared to the DML, the MAP estimation~\eqref{prob:MMVL20_original} can be equivalently reformulated as a MISDP problem due to the additional Tikhonov regularization, and then, its global optimum can be conveniently obtained by a state-of-the-art MISDP solver.

Problem~\eqref{prob:MMVL20_original} is a generalization of the $\ell_0$-norm constrained LS regression problem for a single measurement, as considered by Pilanci et al.~\cite{pilanciSparseLearningBoolean2015}, to the MMV.
Although the objective function in~\eqref{prob:MMVL20_original} can be rewritten as a regularized LS problem for a single measurement vector by vectorizing the matrices $\vect{Y}$ and $\vect{X}$, the coupling among the columns of $\vect{X}$ introduced by the $\ell_{2,0}$-norm constraint cannot be easily addressed by the vectorization $ \vec(\vect{X}) $.
Consequently, Pilanci's MISDP reformulation for the SMV case, as presented in~\cite{pilanciSparseLearningBoolean2015}, cannot be directly applied to problem~\eqref{prob:MMVL20_original}.
In this paper, we provide a nontrivial extension of Pilanci's MISDP reformulation to the MMV case by directly working with the matrix variable $\vect{X}$ and exploiting the row-sparsity structure.
%Thus, a similar MISDP reformulation can be obtained for problem~\eqref{prob:MMVL20_original} by following the line of analysis similar to that in~\cite{pilanciSparseLearningBoolean2015}.
%The additional Tikhonov regularization term $\rho \norm{\vect{X}}_\Frob^2$ with $\rho > 0$ is required for the MISDP reformulation, which will be clarified afterward.
%Moreover, problem~\eqref{prob:MMVL20_original}, as a penalized LS problem, can also be interpreted as a discrete variant of the MAP estimator with a Gaussian prior, which we discuss in detail in Section~\ref{sec:bayesian}.

The parameter $\rho$ is set via~\eqref{eq:rho} if the source power is known and satisfies the assumption in~\eqref{eq:prior}.
Otherwise, if training data are available, a suitable $\rho$ may be obtained through cross-validation or, more efficiently, via the algorithm unrolling procedure~\cite{mongaAlgorithmUnrollingInterpretable2021}.
Also, the parameter $\rho$ has a significant influence on the running time of the two solution approaches for problem~\eqref{prob:MMVL20_original} introduced in Section~\ref{sec:solution}. As observed, a larger $\rho$ leads to a better-conditioned problem, which can generally be solved faster. Therefore, without prior information, one may choose $\rho$ to be as small as possible under a given running time limit, to achieve a good approximation of the DML estimator with the efficient solution approaches presented in Section~\ref{sec:solution}.

In the following, we present a simplified derivation of the MISDP reformulation for the MMV case, which, unlike that in~\cite{pilanciSparseLearningBoolean2015}, does not involve the dual problem constructed with the Legendre-Fenchel conjugate.
Instead, we justify the equivalence of the MISDP reformulation by the concentration with respect to the nuisance parameters in the primal domain.
First, by introducing additional binary variables~$\bu \in \{0,1\}^K$, the original $\ell_{2,0}$-norm constrained problem in~\eqref{prob:MMVL20_original} is equivalently represented as the lifted problem
\begin{align}
	\min_{\substack{\bu \in \{0,1\}^K\\ \mat{u}^\Trans \mat{1} \leq L}} \min_{\bX \in \Compl^{K \times N}} \
	\norm{\bA \bD(\bu) \bX - \bY}_\Frob^2 +
	\rho \norm{\bX}_\Frob^2. \label{prob:MMVL20IP1}
\end{align}
%	where $\vect{1}$ is a vector with all entries equal to one.
%Here, we use the notation~$X_{\bcdot j}$ to denote the~$j$-th column of~$X$ and~$X_{i\bcdot}^T$ to denote the~$i$-th row of~$X$.
The matrix $\mat{D} (\mat{u})$ is a diagonal matrix with $\mat{u}$ on its main diagonal, which determines the active directions in the dictionary $\vect{A}$, i.e., the directions with nonzero source signals.
Note that $ \bD(\bu) $ is not required in the regularization in~\eqref{prob:MMVL20IP1} because the rows of $ \bX $ that are not selected by $ \mat{D}(\bu) $ are not involved in the data fitting term and, hence, will be enforced to be all-zero by the minimization of $ \norm{\bX}_\Frob^2 $.
%\footnote{Note that the MAP estimation problem in~\eqref{prob:MAP} is often solved by grid search in practice. By adding $\vect{D}(\vect{u})$ into the Tikhonov regularization in~\eqref{prob:MMVL20IP1}, problem~\eqref{prob:MMVL20IP1} can also be interpreted as a discretized version of the MAP estimation problem in~\eqref{prob:MAP}. That is, problem~\eqref{prob:MMVL20IP1} is equivalent to solving the MAP estimation problem in~\eqref{prob:MAP} via a brute-force search over the grid $\{\nu_k\}_{k=1}^K$ in~\eqref{eq:on-grid}.}
Like the MAP estimation~\eqref{prob:MAP}, problem~\eqref{prob:MMVL20IP1} is concentrated with respect to $\vect{X}$ and then reformulated by the matrix inversion lemma as the following integer program (IP)
\begin{equation}\label{prob:MMVL20IP2}
	\min_{\substack{\bu \in \{0,1\}^K, \, \mat{u}^\Trans \mat{1} \leq L}} \, \tr \left( \bY^\Herm (\tfrac{1}{\rho} \bA \bD(\bu) \bA^\Herm + \bI_M)^{-1} \bY\right).
\end{equation}

Next, by applying the same SDP reformulation technique as in~\cite{steffensCompactFormulationEll2018,pilanciSparseLearningBoolean2015}, the integer program in~\eqref{prob:MMVL20IP2} can be further written as the MISDP problem with a slack variable $ \bT$:
\begin{subequations}\label{prob:MMVL20MISDP}
	\begin{align}
		\min_{\bu \in \{0,1\}^K,\bT \in \Sym_+^N} \quad & \tr (\bT) \\
		\st \quad
		&
		\begin{bmatrix}
			\tfrac{1}{\rho} \bA \bD(\bu) \bA^\Herm + \bI_M & \bY \\
			\bY^\Herm & \bT
		\end{bmatrix} \succeq 0, \label{eq:MMVL20MISDP_constraint} \\
		& \mat{u}^\Trans \mat{1} \leq L.
	\end{align}
\end{subequations}
%	where $\Sym_+^N$ is the set of $N \times N$ positive semidefinite Hermitian matrices.
The positive semidefiniteness of $\vect{T}$ is enforced by the PSD constraint~\eqref{eq:MMVL20MISDP_constraint}.
The equivalence of~\eqref{prob:MMVL20IP2} and~\eqref{prob:MMVL20MISDP} can be shown as follows.
Since $\tfrac{1}{\rho} \bA \bD(\bu) \bA^\Herm + \bI_M$ is positive definite, by the Schur complement formula, the constraint~\eqref{eq:MMVL20MISDP_constraint} is equivalent to
% \begin{equation*}
$ \bT \succeq \bY^\Herm (\tfrac{1}{\rho} \bA \bD(\bu) \bA^\Herm + \bI_M)^{-1} \bY $~\cite{vandenbergheSemidefiniteProgramming1996}.
% \end{equation*}
Therefore, for every given $\mat{u}$, the minimum of $\tr(\mat{T})$ in~\eqref{prob:MMVL20MISDP} is achieved at
%\begin{equation}\label{key}
$\bT = \bY^\Herm (\tfrac{1}{\rho} \bA \bD(\bu) \bA^\Herm + \bI_M)^{-1} \bY$.
%\end{equation}
The above argument also exhibits the fact that the solution of the MISDP problem~\eqref{prob:MMVL20MISDP} is completely determined as long as the optimal solution of the binary variable $\vect{u}$ is given.

% Apart from the statistical interpretation of the Tikhonov regularization in~\eqref{prob:MAP}, the necessity of having a strictly positive regularization parameter $\rho$ is also revealed by the above MISDP reformulation. First, the Tikhonov regularization enables the reformulation to a MISDP, which has been extensively studied in the literature and can be efficiently solved to global optimality by branch-and-bound solvers, e.g., SCIP-SDP~\cite{gallyFrameworkSolvingMixedinteger2018,matterPresolvingMixedIntegerSemidefinite2023,hojnyHandlingSymmetriesMixedInteger2023}.
Beyond its statistical role, the Tikhonov regularization is also necessary for the above MISDP reformulation, which enables an efficient solution approach to global optimality by branch-and-bound solvers, e.g., SCIP-SDP~\cite{gallyFrameworkSolvingMixedinteger2018,matterPresolvingMixedIntegerSemidefinite2023,hojnyHandlingSymmetriesMixedInteger2023}.
A customized branch-and-bound method can similarly be designed for the sparse formulation of DML, i.e., problem~\eqref{prob:MMVL20IP1} with $\rho = 0$.
% the lack of Tikhonov regularization leads to the following difficulty in the branch-and-bound procedure. At each branch-and-bound node, a continuous subproblem where a part of the binary variables are fixed and the other part are relaxed to continuous needs to be solved. However, in contrast to the proposed regularized formulation, the concentrated sparse DML cannot be easily solved even when $ \vect{u} $ takes continuous values. % due to the nonconvexity.
However, without the regularization, the continuous relaxations encountered at each branch-and-bound node remain nonconvex and computationally difficult to solve.

%However, one drawback of the MISDP formulation in~\eqref{prob:MMVL20MISDP} is the unscalability with respect to the number of snapshots $ N $ since
The dimension of the semidefinite constraint~\eqref{eq:MMVL20MISDP_constraint} is proportional to the number of snapshots $ N $, which becomes computationally demanding for large $ N $.
Following a similar line of analysis as in SPARROW~\cite{steffensCompactFormulationEll2018}, we derive an equivalent MISDP formulation that scales better with respect to $N$.
%whose dimension is independent of $ N $.
The objective function in~\eqref{prob:MMVL20IP2} depends only on the sample covariance matrix $ \widehat{\bR} = \tfrac{1}{N} \bY \bY^\Herm $ and can be rewritten as
\begin{equation}\label{eq:MMVL20IP3}
	\begin{aligned}
		&\quad \tr \left( \bY^\Herm (\tfrac{1}{\rho} \bA \bD(\bu) \bA^\Herm + \bI_M)^{-1} \bY\right)\\
		&= N \tr \left( (\tfrac{1}{\rho} \bA \bD(\bu) \bA^\Herm + \bI_M)^{-1} \widehat{\bR} \right) \\
		&= \tr \left( \widehat{\bY}^\Herm (\tfrac{1}{\rho} \bA \bD(\bu) \bA^\Herm + \bI_M)^{-1} \widehat{\bY}\right),
	\end{aligned}
\end{equation}
where $ \widehat{\bY} = \sqrt{N} \cdot \widehat{\bR}^{\frac{1}{2}} \in \Compl^{M \times M} $.
%Consequently, an equivalent MISDP formulation can be obtained by replacing $\vect{Y}$ in~\eqref{prob:MMVL20MISDP} by $\widehat{\bY}$, which is preferable in the oversampled case, i.e., $N > M$, since the dimension of its semidefinite constraint is independent of the number of snapshots $N$.
It leads to the following equivalent MISDP formulation:
\begin{subequations}
	\label{prob:MMVL20MISDP2}
	\begin{align}
		\underset{\bu \in \{0,1\}^K, \bT \in \Sym_+^M}{\min}\quad
		& \tr(\bT) \\
		\text{s.t.}\quad
		&
		\begin{bmatrix}
			\tfrac{1}{\rho} \bA \bD(\bu) \bA^\Herm + \bI_M & \widehat{\bY} \\
			\widehat{\bY}^\Herm & \bT
		\end{bmatrix} \succeq 0, \label{eq:MMVL20MISDP2_constraint} \\
		& \mat{u}^\Trans \mat{1} \leq L.
	\end{align}
\end{subequations}
In contrast to~\eqref{eq:MMVL20MISDP_constraint}, the dimension of the semidefinite constraint~\eqref{eq:MMVL20MISDP2_constraint} is independent of $ N $. In summary, either formulation~\eqref{prob:MMVL20MISDP} or~\eqref{prob:MMVL20MISDP2} can be used to solve the $ \ell_{2,0} $-norm constrained problem in~\eqref{prob:MMVL20_original}, depending on the number of snapshots $ N $. Specifically, \eqref{prob:MMVL20MISDP} is preferable in the undersampled case, i.e., $ N \leq M $, and~\eqref{prob:MMVL20MISDP2} is preferable otherwise.

\section{Solution Approaches for the MISDP}
\label{sec:solution}

The MISDP implementation~\eqref{prob:MMVL20MISDP} or~\eqref{prob:MMVL20MISDP2} of the MMV problem can be directly solved by a general purpose MISDP solver such as SCIP-SDP~\cite{gallyFrameworkSolvingMixedinteger2018,matterPresolvingMixedIntegerSemidefinite2023,hojnyHandlingSymmetriesMixedInteger2023}. The SCIP-SDP solver provides two efficient approaches based on the widely used branch-and-bound method and cutting-plane method.
Both approaches admit an improved scalability for large problems compared to the simple brute-force search on the integers.
Compared to this setting, the branch-and-bound method, even if terminated early, provides an optimality assessment of the solution by the gap between the lower and upper bounds of the optimal value.
Both approaches require iteratively either solving continuous SDP relaxations or the computation of an eigendecomposition for the construction of cutting planes~\cite{gallyFrameworkSolvingMixedinteger2018}.
Hence, the SCIP-SDP solver may become computationally demanding with the increase of the problem dimension due to the increase in the number of required nodes in the branch-and-bound tree and the number of cutting planes.
% and the computational complexity of each subproblem.

In contrast to solving the MISDP exactly, a low-complexity method for obtaining a good approximate solution in the case with a single measurement vector is presented in~\cite{pilanciSparseLearningBoolean2015} based on the randomized rounding (RR) technique. The method can be generalized to the MMV case and it consists of two main steps, namely, interval relaxation and randomized rounding, which is outlined in Algorithm~\ref{algo1}.
\begin{algorithm}[h]
	\caption{The Randomized Rounding (RR) Algorithm for the MISDP Problem~\eqref{prob:MMVL20MISDP} or~\eqref{prob:MMVL20MISDP2}.}
	\label{algo1}
	\begin{enumerate}[leftmargin=*,rightmargin=1.5em,label={Step \arabic*}]
		\item \label{step1} \textit{Interval Relaxation:} Solve the convex relaxation of the MISDP~\eqref{prob:MMVL20MISDP} or~\eqref{prob:MMVL20MISDP2} that replaces the Boolean hypercube $\{0,1\}^K$ by its convex hull, the unit hypercube $[0,1]^K$. Let $\widehat{\vect{u}} \in [0,1]^K$ be the optimal solution of the relaxed problem.
		\item \label{step2} \textit{Randomized Rounding:} Given the approximate solution $\widehat{\vect{u}}$, randomly generate $T > 0$ binary solutions $\widetilde{\vect{u}} \in \{0,1\}^K$ where each entry $\tilde{u}_k$ independently follows the Bernoulli distribution
			%	\begin{equation}\label{key}
			$\Pr [\tilde{u}_k = 1] = \hat{u}_k$  and $\Pr [\tilde{u}_k = 0] = 1-\hat{u}_k$
			%	\end{equation}

			for $k = 1,\ldots,K$. An infeasible binary solution with $\widetilde{\vect{u}}^\Trans \vect{1} > L$ is afterward projected to the feasible set by retaining only the $L$ ones in $\widetilde{\vect{u}}$ with the largest values in the fractional solution $\widehat{\vect{u}}$. Then, the solution $\widetilde{\vect{u}}^\star$ with the smallest objective function value, which can be calculated according to the IP formulation in~\eqref{prob:MMVL20IP2}, among all feasible solutions is chosen to be the approximate solution.
	\end{enumerate}
\end{algorithm}

The interval relaxation in Step 1 is a continuous SDP and can be efficiently solved by a generic interior-point solver such as MOSEK~\cite{mosekapsMOSEKOptimizationToolbox2023}.
To further reduce the complexity of Step 1 in the case of large dimensions, one may consider an attractive alternative that
% , instead of solving the interval relaxation of the MISDP formulation in~\eqref{prob:MMVL20MISDP} with interior-point solvers,
employs projected first-order and quasi-Newton methods~\cite{schmidtOptimizingCostlyFunctions2009}, or primal-dual methods such as ADMM~\cite{boydDistributedOptimizationStatistical2011} to solve the interval relaxation of the equivalent IP formulation in~\eqref{prob:MMVL20IP2}. When the optimal solution $\widehat{\vect{u}}$ of the interval relaxation is binary, then it is also the optimal solution of the original MISDP problem.
Additional analyses of the conditions for the interval relaxation to have a binary solution are provided in~\cite{pilanciSparseLearningBoolean2015} in the SMV case with random dictionaries.

Otherwise, if the solution $\widehat{\vect{u}}$ is not binary, the randomized rounding technique is employed to efficiently search for a good binary solution. The basic randomized rounding procedure, which is used in~\cite{pilanciSparseLearningBoolean2015}, searches in a set of $T>0$ randomly generated independent and identically distributed (i.i.d.) binary solutions $\widetilde{\vect{u}} \in \{0,1\}^K$ where each entry $\tilde{u}_k$ independently follows the Bernoulli distribution with the success probability
\begin{equation} \label{eq:binaryProb}
	\Pr [\tilde{u}_k = 1] = \hat{u}_k.
\end{equation}

The motivation for using the fractional solution as the probability is as follows. First, the fractional solution is restricted to be in the interval $[0,1]$. Second, the solution of the continuous relaxation is expected to be close to the solution of the original integer program, which relies on the tightness of the relaxation. In particular, several bounds on the disparity between the optimal value of the interval relaxation and that of the original problem have been established for the class of binary integer linear programs~\cite{raghavanRandomizedRoundingTechnique1987,motwaniRandomizedAlgorithms1995,williamsonDesignApproximationAlgorithms2011}. Although a theoretical guarantee on the tightness of the interval relaxation for the considered MISDP problem has not been derived, the numerical simulations in Section~\ref{sec:results} show that randomized rounding is capable of finding good approximate solutions for the considered MISDP problem with reasonable choices of $T$.

%Randomized rounding in Step 2 is a classical technique for converting fractional solutions into binary solutions with provable approximation guarantees~\cite{raghavanRandomizedRoundingTechnique1987,motwaniRandomizedAlgorithms1995,williamsonDesignApproximationAlgorithms2011}. Since the fractional solution $\widehat{\vect{u}}$ satisfies $\widehat{\vect{u}}^\Trans \vect{1} \leq L$,
The cardinality of a binary solution $\widetilde{\vect{u}}$ generated by the distribution~\eqref{eq:binaryProb} follows a Poisson binomial distribution with the expectation
%\begin{equation}\label{key}
$\Expect [\widetilde{\vect{u}}^\Trans \vect{1}] = \widehat{\vect{u}}^\Trans \vect{1} \leq L$.
%\end{equation}
If the expected sparsity level $L$ is consistent with the ground truth, the interval relaxation, as well as the original integer program, is expected to attain the optimum at the boundary of the constraint set $\{\vect{u} \mid \vect{u}^\Trans \vect{1} \leq L\}$.
In this case, i.e., when $\widehat{\vect{u}}^\Trans \vect{1} = L$, a drawback of the basic randomized rounding is that the probability of generating an infeasible binary solution, i.e., $\Pr [\widetilde{\vect{u}}^\Trans \vect{1} > L]$, is still high.
%which ensures the high probability of the existence of feasible solutions among the randomly generated binary solutions for reasonable choices of $T$.
%Although Pilanci's method can be trivially generalized to the MMV case as considered here, the interval relaxation becomes less tight when the number of snapshots increases, as demonstrated by the simulation results in Section~\ref{sec:results}.
One may increase $T$ to explore a sufficient number of feasible binary solutions. Alternatively, an extended version of randomized rounding with scaling is proposed in~\cite{raghavanRandomizedRoundingTechnique1987} for reducing the failure probability $\Pr [\widetilde{\vect{u}}^\Trans \vect{1} > L]$. In short, it uses a down-scaled version $(1-\delta) \widehat{\vect{u}}$, with $\delta \in (0,1)$, of the fractional solution, rather than directly $\widehat{\vect{u}}$, as the probabilities for the generation of binary solutions. Then the expected cardinality of the generated binary solutions does not exceed $(1-\delta) L$. A required upper bound on the failure probability $\Pr [\widetilde{\vect{u}}^\Trans \vect{1} > L]$ can be satisfied by properly choosing $\delta$. Nevertheless, since random samples concentrate on the expectation, more binary solutions in the interior of the constraint set $\{\vect{u} \mid \vect{u}^\Trans \vect{1} \leq L\}$ are explored by down-scaling the probabilities.

As we consider the case where the expected sparsity level $L$ is consistent with the ground truth, it is instead beneficial to test more binary solutions at the boundary, i.e., with a cardinality of exactly $L$.
To this end, we propose a different extended version of randomized rounding with a subsequent projection step. As described in Algorithm~\ref{algo1}, instead of discarding the generated binary solutions with cardinality larger than $L$, we project each infeasible binary solution to the constraint set by preserving only $L$ ones with the largest probabilities, i.e., the values in the fractional solution $\widehat{\vect{u}}$.
Consistent with the above analysis, the proposed RR with projection often finds a binary solution that provides a lower objective function value even with fewer randomly explored binary solutions compared to the basic RR and RR with scaling.

The IP formulation~\eqref{prob:MMVL20IP2} established on the sparse signal model~\eqref{eq:sparseModel} is equivalent to solving the concentrated MAP estimation problem~\eqref{prob:concentratedMAP2} via a brute-force search over the sampling grid $\{\nu_k\}_{k=1}^K$.
Compared to the objective function in~\eqref{prob:concentratedMAP2}, the equivalent expression in~\eqref{prob:concentratedMAP1} admits a lower computational complexity since the number of sources $L$ is generally much smaller than the array size $M$. Therefore, the evaluation of the objective function in~\eqref{prob:MMVL20IP2} at a randomly generated binary solution $\widetilde{\vect{u}}$ in the randomized rounding can be more efficiently performed according to the objective function~\eqref{prob:concentratedMAP1}, where the steering matrix $\vect{A}(\vect{\mu})$ is determined by the columns in the dictionary $\vect{A}(\vect{\nu})$ selected by $\widetilde{\vect{u}}$.

%Note that the same relaxation technique is used to construct the continuous relaxed subproblems in the branch-and-bound approach of SCIP-SDP.
We remark that a basic randomized rounding procedure without the subsequent projection step in Algorithm~\ref{algo1} is employed internally by SCIP-SDP to find good integer solutions for the estimation of the upper bounds of the optimal value in the branch-and-bound approach.
%Like other branch-and-bound methods, SCIP-SDP is typically observed to quickly find an optimal or nearly optimal solution but it requires much more time to verify the optimality of the obtained solution via lower bounds.
Similar to other branch-and-bound methods, SCIP-SDP typically finds an optimal or near-optimal solution rapidly. However, it requires significantly more time to verify the optimality of this solution through lower bound computations.
This observation also motivates the idea of using the RR Algorithm~\ref{algo1} for obtaining a good approximate solution at a low computational cost.

The on-grid assumption~\eqref{eq:on-grid} is typically not fulfilled in practice due to the finite grid size, which results in spectral leakage effects and basis mismatch~\cite{chiSensitivityBasisMismatch2011,hermanGeneralDeviantsAnalysis2010} in the reconstructed signal.
To mitigate the grid mismatch, an additional gridless local search step can be performed to find a local optimum of the gridless DML estimation in~\eqref{prob:concentratedDML} or the gridless MAP estimation~\eqref{prob:concentratedMAP1}, starting from the frequencies recovered by the grid-based method. Examples of gridless refinement methods include gradient methods~\cite{parkAtomconstrainedGridlessDOA2024,gerstoftAtomconstrainedMaximumLikelihood2025} and the alternating projection algorithm~\cite{ziskindMaximumLikelihoodLocalization1988}.
Since it finds only a local optimum, it is not guaranteed to always improve the estimation quality, especially if the frequencies recovered by the grid-based method are very poor.

%{\red An additional step for further improving the solution of Pilanci's method is also evaluated in the simulation, which employs the \texttt{fmincon} solver provided by MATLAB with the interior-point algorithm to find a local optimum for the DML estimation starting from the frequencies recovered by Pilanci's method.}

\section{Relation to the $\ell_{2,1}$-Norm Minimization}
\label{sec:SPARROW}
%In contrast to our proposed method, the SPARROW method in~\cite{steffensCompactFormulationEll2018} utilizes the convex $\ell_{2,1}$-norm regularized formulation~\cite{malioutovSparseSignalReconstruction2005}
The $\ell_1$-norm is widely used as a convex approximation of the $\ell_0$-norm for obtaining computationally more tractable problems~\cite{tibshiraniRegressionShrinkageSelection1996,chenAtomicDecompositionBasis2001}.
In the MMV case, the $\ell_{2,1}$-norm minimization problem~\cite{malioutovSparseSignalReconstruction2005}
\begin{equation}\label{prob:SPARROW_original}
	\min_{\vect{X} \in \Compl^{K \times N}} \ \tfrac{1}{2} \norm{\vect{AX} - \vect{Y}}_\Frob^2 + \lambda \sqrt{N} \norm{\vect{X}}_{2,1}
\end{equation}
is typically considered as a generalization of the classic $\ell_1$-norm minimization problem,
where $\lambda > 0$ is a regularization parameter.
%Problem~\eqref{prob:SPARROW_original} is a generalization of the widely studied $\ell_1$-norm minimization problem~\cite{tibshiraniRegressionShrinkageSelection1996,chenAtomicDecompositionBasis2001} to the MMV case.
To address problem~\eqref{prob:SPARROW_original}, the SPARROW method~\cite{steffensCompactFormulationEll2018} utilizes the equivalent convex reformulation
\begin{equation}\label{prob:SPARROW_reform}
	\underset{\vect{s} \in \Real_+^K}{\min} \ \tr \left( (\vect{AD}(\vect{s}) \vect{A}^\Herm + \lambda \vect{I}_M)^{-1} \widehat{\vect{R}} \right) + \vect{s}^\Trans \vect{1}.
\end{equation}
The optimal solutions $\widehat{\vect{X}} = [\widehat{\vect{x}}_1, \ldots, \widehat{\vect{x}}_K]^\Trans$ and $\widehat{\vect{s}}$ for~\eqref{prob:SPARROW_original} and~\eqref{prob:SPARROW_reform}, respectively, are related by
%\begin{equation}\label{key}
$\hat{s}_k = \tfrac{1}{\sqrt{N}}\norm{\widehat{\vect{x}}_k}_2$
%\end{equation}
for $k=1,\ldots,K$.
Similarly, problem~\eqref{prob:SPARROW_reform} can be further reformulated as an SDP problem and solved by a generic SDP solver.
Alternatively, a coordinate descent algorithm is also devised in~\cite{steffensCompactFormulationEll2018} for problem~\eqref{prob:SPARROW_reform}, which is more scalable in the case with a large number of sensors $M$.
By the expression in~\eqref{eq:MMVL20IP3}, the integer program~\eqref{prob:MMVL20IP2} can be rewritten as
\begin{equation}\label{prob:MMVL20IP4}
	\min_{\substack{\bu \in \{0,1\}^K, \, \mat{u}^\Trans \mat{1} \leq L}} \, \tr \left( \left( \bA \bD(\bu) \bA^\Herm + \rho \bI_M\right)^{-1} \widehat{\vect{R}}\right),
\end{equation}
where some constant factors are discarded.
Thus, it is concluded that problem~\eqref{prob:SPARROW_reform} is equivalent to a Lagrangian of a convex continuous relaxation of the integer program~\eqref{prob:MMVL20IP4} where, in contrast to the interval relaxation used by the randomized rounding Algorithm~\ref{algo1}, the binary variables are further relaxed to be nonnegative.
As shown by the simulations in Section~\ref{sec:results}, the continuous relaxation and Lagrangian relaxation lead to a degradation of the estimation quality, compared to our proposed method.
%Moreover, the solution $\widehat{\vect{s}}$ is not as sparse as the desired number of sources $L$. Then the frequencies associated with the $L$ peaks in $\widehat{\vect{s}}$ with the largest prominence are chosen to be the estimate.

\section{Generalization to the Sparse MAP Estimation with Nonuniform Source Variances}
\label{sec:generalization}

In Section~\ref{sec:problem}, we developed a MISDP reformulation for the sparse MAP estimation~\eqref{prob:MMVL20_original} with the uniform prior~\eqref{eq:prior_sparse} where a priori the sources in all directions have the same variance. In this section, we consider the sparse MAP estimation problem with nonuniform source variances and briefly introduce a generalization of the MISDP-based method in Section~\ref{sec:problem}.

Consider the sparse model in~\eqref{eq:sparseModel} constructed by sampling the FOV. The waveform matrix $\vect{X}$ is similarly assumed to be row-sparse and follow a spatial-temporal i.i.d. zero-mean complex Gaussian distribution. However, unlike the uniform prior assumption in~\eqref{eq:prior_sparse}, the source variances are assumed to vary across the DOAs.
Specifically, if there exists a source in the sampled direction with the spatial frequency $\nu_k$, then the entries $x_{k,n}$ are assumed to follow the distribution
\begin{equation} \label{eq:prior_nonuniform}
	x_{k,n} \sim \mathcal{CN} (0,\gamma_k)
\end{equation}
for $n = 1,\ldots,N$, where $\gamma_k>0$ denotes the given potential source variance in the DOA with the spatial frequency $\nu_k$.
The vector $\vect{\gamma} = [\gamma_1,\ldots,\gamma_K]^\Trans$ can be considered as the sampled spatial source power spectrum over the FOV, which may be estimated a priori by, e.g., the conventional beamforming or the SBL method.
Define the regularization parameters
\begin{equation}
	\rho_k = {\sigma^2}/{\gamma_k}
\end{equation}
for $k = 1,\ldots,K$ and summarize them in the vector $\vect{\rho} = [\rho_1,\ldots,\rho_K]^\Trans$.
Then, using the same considerations as in Section~\ref{sec:problem}, we can write the sparse MAP estimation problem corresponding to the nonuniform prior in~\eqref{eq:prior_nonuniform} as the following sparse LS problem with a row-wise weighted Tikhonov regularization
\begin{equation} \label{prob:MMVL20IP_rowwise1}
	\min_{\substack{\bu \in \{0,1\}^K\\ \mat{u}^\Trans \mat{1} \leq L}} \min_{\bX \in \Compl^{K \times N}} \
	\norm{\bA \bD(\bu) \bX - \bY}_\Frob^2 + \norm{\vect{D}(\sqrt{\vect{\rho}}) \bX}_\Frob^2,
\end{equation}
where the square root operation is performed elementwise and $\rho_k$ represents the regularization weight for the squared Euclidean norm of the $k$th row of $\vect{X}$.
Similar to problem~\eqref{prob:MMVL20IP1}, the matrix $ \bD(\bu) $ is not required in the regularization in~\eqref{prob:MMVL20IP_rowwise1} because the rows of $ \bX $ that are not selected by $ \mat{D}(\bu) $ are not involved in the data fitting term and will be enforced to be all-zero by the minimization of the row norm.

Likewise, problem~\eqref{prob:MMVL20IP_rowwise1} can be concentrated with respect to $\vect{X}$ and then reformulated by the matrix inversion lemma as the following integer program
\begin{equation}\label{prob:MMVL20IP_rowwise2}
	\min_{\substack{\bu \in \{0,1\}^K, \, \mat{u}^\Trans \mat{1} \leq L}} \, \tr \left( \bY^\Herm (\bA \bD(\bu \oslash \vect{\rho}) \bA^\Herm + \bI_M)^{-1} \bY\right),
\end{equation}
where $\oslash$ denotes the elementwise division.
The integer program in~\eqref{prob:MMVL20IP_rowwise2} can be further written as MISDP problems similar to~\eqref{prob:MMVL20MISDP} and~\eqref{prob:MMVL20MISDP2}.
% by applying the same reformulation technique as in Section~\ref{sec:problem}.
Likewise, the MISDP reformulation is solved exactly by the SCIP-SDP solver or approximately by the randomized rounding Algorithm~\ref{algo1}, as discussed in Section~\ref{sec:solution}. The details of the generalization of the MISDP reformulation and the solution approach for the sparse MAP estimation problem in~\eqref{prob:MMVL20IP_rowwise1} with nonuniform source variances are omitted due to space limitations.

%{\red summarize how to choose the regularization parameter, uniform or rowwise}

\section{Simulation Results}
\label{sec:results}
We conduct numerical experiments on synthetic data to evaluate and analyze the performance of the proposed method, including both solution approaches introduced in Section~\ref{sec:solution} for the MISDP reformulation. Specifically, the SCIP-SDP solver~\cite{gallyFrameworkSolvingMixedinteger2018,matterPresolvingMixedIntegerSemidefinite2023,hojnyHandlingSymmetriesMixedInteger2023} of version 4.1.0 is used and the nonlinear branch-and-bound approach is chosen in the SCIP-SDP solver, where the relaxed continuous SDP subproblems are solved by MOSEK~\cite{mosekapsMOSEKOptimizationToolbox2023} of version 10.0.38.\footnote{The source code of SCIP-SDP and an interface for MATLAB can be downloaded from \url{https://wwwopt.mathematik.tu-darmstadt.de/scipsdp} and \url{https://github.com/scipopt/SCIP-SDP}. In the experiments, the SCIP-SDP solver is called from MATLAB through the provided interface.}
The additional gridless local search in Section~\ref{sec:solution} is achieved by the \texttt{fmincon} solver with the interior-point algorithm provided by MATLAB.
The continuous SDP problem in the randomized rounding Algorithm~\ref{algo1} is modeled by using CVX~\cite{grantCVXMatlabSoftware2020,grantGraphImplementationsNonsmooth2008} and solved by MOSEK.
The estimation error of the proposed method is compared to the stochastic Cram\'er-Rao Bound (CRB)~\cite{stoicaStochasticCRBArray2001} and the estimation error of several widely used approaches for DOA estimation, namely, MUSIC~\cite{schmidtMultipleEmitterLocation1986}, root-MUSIC~\cite{barabellImprovingResolutionPerformance1983}, the SPARROW method with coordinate descent implementation, the sparse Bayesian learning (SBL) method in~\cite{gerstoftMultisnapshotSparseBayesian2016}, and the DML estimator.
%which is considered to be statistically optimal.
The DML estimator is obtained via a brute-force search over the same grid as in~\eqref{eq:sparseModel}, which is equivalent to the solution of problem~\eqref{prob:MMVL20_original} with no regularization.

The SBL method in~\cite{gerstoftMultisnapshotSparseBayesian2016} also employs an uncorrelated Gaussian prior assumption. However, in the SBL method, the source variances are not assumed to be known but estimated via type-II maximum likelihood.
Whereas SBL theoretically recovers source support without knowing the number of sources $L$, the effect of noise and the finite number of snapshots yield nonsparse variance estimates in practice. Thus, similar to other spectral methods such as MUSIC, the $L$ largest peaks in the spectrum of source variances estimated by SBL are chosen to be the estimated frequencies.
% Ideally, SBL does not require the knowledge of the number of sources since the estimated source variances directly reveal the number of sources and their locations. In practice, due to the effect of noise and the finite number of snapshots, the estimated source variances are typically nonsparse.

The results are averaged over $N_R = 1000$ Monte-Carlo trials unless otherwise noted.
In particular, the quality of the estimated spatial frequencies $\widehat{\vect{\mu}}(n) = [\hat{\mu}_1(n), \ldots, \hat{\mu}_L(n)]^\Trans$ for $n=1,\ldots,N_R$ is measured by the root-mean-square error (RMSE) with respect to the ground-truth $\vect{\mu}$ defined as
\begin{equation}
	\text{RMSE} (\widehat{\bmu}) =
	% \begin{matrix}
		\sqrt{\frac{1}{L N_R} \sum_{n=1}^{N_R} \sum_{l=1}^{L} \abs{\hat{\mu}_l (n) - \mu_l}_\text{wa}^2}
	% \end{matrix},
\end{equation}
where $ \abs{\mu_1 - \mu_2}_\text{wa} = \min_{k \in \mathbb{Z}} \abs{\mu_1 - \mu_2 + 2 k \pi} $ denotes the wrap-around distance between two frequencies $\mu_1$ and $\mu_2$.
All experiments were conducted on a Linux PC with an Intel Core i7-7700 CPU and 32 GB RAM running MATLAB 2024a.

We consider a ULA of $M=8$ sensors with half-wavelength inter-element spacing,
and the dictionary $\vect{A}$ is constructed with $K=100$ grid points with frequencies uniformly sampled in $[-\pi, \pi)$. The SNR is calculated as $\text{SNR} = 1/ \sigma^2$.

Some algorithmic parameters are set as follows. The regularization parameter $\rho$ in~\eqref{prob:MMVL20_original} for our proposed method is chosen according to the rule in~\eqref{eq:rho} and the parameter $\lambda$ in~\eqref{prob:SPARROW_original} for the SPARROW method is selected by the heuristic rule
\begin{equation}\label{eq:lambda}
	\lambda = \sqrt{\sigma^2 M \log M}
\end{equation}
as recommended in~\cite{bhaskarAtomicNormDenoising2013,steffensCompactFormulationEll2018}.
The number of randomly generated binary solutions in the RR algorithm is $T = 10^4$ for $L=3$ sources and $T = 10^5$ for $L=5$ as the size of the binary feasible set of problem~\eqref{prob:MMVL20MISDP} increases exponentially with $L$.
If not specified, the two equivalent MISDP reformulations~\eqref{prob:MMVL20MISDP} and~\eqref{prob:MMVL20MISDP2} are chosen according to the discussion in Section~\ref{sec:problem} to achieve a lower computation time.

\subsection{Uncorrelated Source Signals}
\label{subsec:uncorrelated}
In the following, we evaluate the performance of the methods through seven numerical experiments under the assumption of uncorrelated source signals, whereas the influence of source correlation is analyzed in Section~\ref{subsec:correlated}.
% We first compare the performance of the methods for uncorrelated source signals that satisfy the prior assumption in~\eqref{eq:prior}.
Specifically, for each Monte-Carlo trial, the true source signals in $\vect{\Psi}$ are randomly generated according to the uncorrelated Gaussian prior in~\eqref{eq:prior}
%, i.e., the signals $\vect{\psi}_n$ in different time instants are i.i.d. samples of the Gaussian distribution $\mathcal{CN} (\vect{0},P_\Psi \vect{I}_L)$,
with the variance $\gamma = 1$.

Experiments~1 and~2 analyze the impact of the SNR and the number of snapshots, respectively, on the estimation accuracy.
Experiment~3 investigates the resolution performance by varying the frequency separation between closely located sources.
In Experiments~4--7, we focus on the scalability and computational efficiency of the proposed MISDP reformulations and the RR algorithm by varying the number of sources, the array size, and the grid size.

\begin{figure}[t]
	\centering
	\ref{legend1}

	\subfigure[\label{fig:varySNR_L=3_noGridless}]{
		\begin{tikzpicture}[inner ysep=1.5pt]
	\def\filepath{results/varySNR_L=3_N=8_uncorr_rmse.csv}
	
	\begin{axis}[%
		at={(0,0)},
		scale only axis,
		enlarge x limits={false},
    xmax=20,
		xlabel style={font=\color{white!15!black}, font=\scriptsize},
		xlabel={SNR (dB)},
%		xlabel shift = -2.5pt,
%		xmode = log,
%		xtick = {0.025,0.05,0.1,0.2,0.4},
%		xticklabels = {0.025,0.05,0.1,0.2,0.4},
		%ymin=0,
%		ymax=1,
		ylabel style={font=\color{white!15!black}, font=\scriptsize},
    ylabel={RMSE (rad)},
%		ylabel shift = -3pt,
		ymode = log,
		axis background/.style={fill=white},
		xmajorgrids,
		ymajorgrids,
		legend style={legend cell align=left, align=left, draw=white, % draw=white!15!black, 
			font=\tiny, row sep=-2pt,
			at={(0,0)}, anchor = south west},
		cycle multi list = {
%			solid,dashed,densely dotted\nextlist
%			[2 of]MATLABcolormarklist
			MATLABcolormarklist
%			MATLABcolorlist
		},
%		legend to name=legend1,
%		legend columns=8,
		]

		\addplot table [x index=0, y=SCIP-SDP] {\filepath};
		\addlegendentry{MISDP via SCIP-SDP};
		
%		\addplot[color=MATLABcolor6, densely dashdotted, mark=triangle, every mark/.append style = {solid}] table [x index=0, y=SCIP-SDP-SVD] {\filepath};
%		\addlegendentry{MISDP-S via SCIP-SDP};
%		
%		\pgfplotsset{cycle list shift=-1};
		
		\addplot table [x index=0, y=Pilanci] {\filepath};
		\addlegendentry{MISDP via IRRR};
		
%		\addplot[color=MATLABcolor7, densely dashdotted, mark=triangle, every mark/.append style = {solid}] table [x index=0, y=Pilanci-SVD] {\filepath};
%		\addlegendentry{MISDP-S via IRRR};
%		
%		\pgfplotsset{cycle list shift=-2};
		
		\addplot table [x index=0, y=SPARROW] {\filepath};
		\addlegendentry{SPARROW};
		
		\addplot table [x index=0, y=SBL] {\filepath};
		\addlegendentry{SBL};
		
		\addplot table [x index=0, y=MUSIC] {\filepath};
		\addlegendentry{MUSIC};
		
		\addplot+[color=TUDa-4a, densely dashdotted] table [x index=0, y=root-MUSIC] {\filepath};
		\addlegendentry{root-MUSIC};
		
		\addplot+[color=TUDa-0c, densely dashdotted,mark=null] table [x index=0, y=DML] {\filepath};
		\addlegendentry{DML};
		
		\addplot[color=black] table [x index=0, y=CRB] {\filepath};
		\addlegendentry{CRB};
		
%		\addplot+[dashed] table [x index=0, y=fmincon] {\filepath};
%		\addlegendentry{DML (\texttt{fmincon})};
		\legend{};
		
	\end{axis}
\end{tikzpicture}
	} \\
	\subfigure[\label{fig:varySNR_L=3_gridlessDML}]{\begin{tikzpicture}[inner ysep=1.5pt]
	\def\filepath{results/varySNR_L=3_N=8_uncorr_fminconDML_rmse.csv}
	
	\begin{axis}[%
		at={(0,0)},
		scale only axis,
		enlarge x limits={false},
    xmax=20,
		xlabel style={font=\color{white!15!black}, font=\scriptsize},
		xlabel={SNR (dB)},
%		xlabel shift = -2.5pt,
%		xmode = log,
%		xtick = {0.025,0.05,0.1,0.2,0.4},
%		xticklabels = {0.025,0.05,0.1,0.2,0.4},
		%ymin=0,
%		ymax=1,
		ylabel style={font=\color{white!15!black}, font=\scriptsize},
		ylabel={RMSE (rad)},
%		ylabel shift = -3pt,
		ymode = log,
		axis background/.style={fill=white},
		xmajorgrids,
		ymajorgrids,
		legend style={legend cell align=left, align=left, draw=white, % draw=white!15!black, 
			font=\tiny, row sep=-2pt,
			at={(0,0)}, anchor = south west},
		cycle multi list = {
%			solid,dashed,densely dotted\nextlist
%			[2 of]MATLABcolormarklist
			MATLABcolormarklist
%			MATLABcolorlist
		},
%		legend to name=legend1,
%		legend columns=8,
		]

		\addplot table [x index=0, y=SCIP-SDP] {\filepath};
		\addlegendentry{MISDP via SCIP-SDP};
		
%		\addplot[color=MATLABcolor6, densely dashdotted, mark=triangle, every mark/.append style = {solid}] table [x index=0, y=SCIP-SDP-SVD] {\filepath};
%		\addlegendentry{MISDP-S via SCIP-SDP};
%		
%		\pgfplotsset{cycle list shift=-1};
		
		\addplot table [x index=0, y=Pilanci] {\filepath};
		\addlegendentry{MISDP via IRRR};
		
%		\addplot[color=MATLABcolor7, densely dashdotted, mark=triangle, every mark/.append style = {solid}] table [x index=0, y=Pilanci-SVD] {\filepath};
%		\addlegendentry{MISDP-S via IRRR};
%		
%		\pgfplotsset{cycle list shift=-2};
		
		\addplot table [x index=0, y=SPARROW] {\filepath};
		\addlegendentry{SPARROW};
		
		\addplot table [x index=0, y=SBL] {\filepath};
		\addlegendentry{SBL};
		
		\addplot table [x index=0, y=MUSIC] {\filepath};
		\addlegendentry{MUSIC};
		
		\addplot+[color=TUDa-4a, densely dashdotted] table [x index=0, y=root-MUSIC] {\filepath};
		\addlegendentry{root-MUSIC};
		
		\addplot+[color=TUDa-0c, densely dashdotted,mark=null] table [x index=0, y=DML] {\filepath};
		\addlegendentry{DML};
		
		\addplot[color=black] table [x index=0, y=CRB] {\filepath};
		\addlegendentry{CRB};
		
%		\addplot+[dashed] table [x index=0, y=fmincon] {\filepath};
%		\addlegendentry{DML (\texttt{fmincon})};
		\legend{};
		
	\end{axis}
\end{tikzpicture}} \\
	\subfigure[\label{fig:varySNR_L=3_gridlessMAP}]{\begin{tikzpicture}[inner ysep=1.5pt]
	\def\filepath{results/varySNR_L=3_N=8_uncorr_fminconMAP_rmse.csv}
	
	\begin{axis}[%
		at={(0,0)},
		scale only axis,
		enlarge x limits={false},
    xmax=20,
		xlabel style={font=\color{white!15!black}, font=\scriptsize},
		xlabel={SNR (dB)},
%		xlabel shift = -2.5pt,
%		xmode = log,
%		xtick = {0.025,0.05,0.1,0.2,0.4},
%		xticklabels = {0.025,0.05,0.1,0.2,0.4},
		%ymin=0,
%		ymax=1,
		ylabel style={font=\color{white!15!black}, font=\scriptsize},
		ylabel={RMSE (rad)},
%		ylabel shift = -3pt,
		ymode = log,
		axis background/.style={fill=white},
		xmajorgrids,
		ymajorgrids,
		legend style={legend cell align=left, align=left, draw=white, % draw=white!15!black, 
			font=\tiny, row sep=-2pt,
			at={(0,0)}, anchor = south west},
		cycle multi list = {
%			solid,dashed,densely dotted\nextlist
%			[2 of]MATLABcolormarklist
			MATLABcolormarklist
%			MATLABcolorlist
		},
%		legend to name=legend1,
%		legend columns=8,
		]

		\addplot table [x index=0, y=SCIP-SDP] {\filepath};
		\addlegendentry{MISDP via SCIP-SDP};
		
%		\addplot[color=MATLABcolor6, densely dashdotted, mark=triangle, every mark/.append style = {solid}] table [x index=0, y=SCIP-SDP-SVD] {\filepath};
%		\addlegendentry{MISDP-S via SCIP-SDP};
%		
%		\pgfplotsset{cycle list shift=-1};
		
		\addplot table [x index=0, y=Pilanci] {\filepath};
		\addlegendentry{MISDP via IRRR};
		
%		\addplot[color=MATLABcolor7, densely dashdotted, mark=triangle, every mark/.append style = {solid}] table [x index=0, y=Pilanci-SVD] {\filepath};
%		\addlegendentry{MISDP-S via IRRR};
%		
%		\pgfplotsset{cycle list shift=-2};
		
		\addplot table [x index=0, y=SPARROW] {\filepath};
		\addlegendentry{SPARROW};
		
		\addplot table [x index=0, y=SBL] {\filepath};
		\addlegendentry{SBL};
		
		\addplot table [x index=0, y=MUSIC] {\filepath};
		\addlegendentry{MUSIC};
		
		\addplot+[color=TUDa-4a, densely dashdotted] table [x index=0, y=root-MUSIC] {\filepath};
		\addlegendentry{root-MUSIC};
		
		\addplot+[color=TUDa-0c, densely dashdotted,mark=null] table [x index=0, y=DML] {\filepath};
		\addlegendentry{DML};
		
		\addplot[color=black] table [x index=0, y=CRB] {\filepath};
		\addlegendentry{CRB};
		
%		\addplot+[dashed] table [x index=0, y=fmincon] {\filepath};
%		\addlegendentry{DML (\texttt{fmincon})};
		\legend{};
		
	\end{axis}
\end{tikzpicture}}
	\caption{RMSE vs. SNR for $L=3$ uncorrelated sources, $M=8$ sensors, $N=8$ snapshots, and $K=100$ grid points, \subref{fig:varySNR_L=3_noGridless} without gridless local search, \subref{fig:varySNR_L=3_gridlessDML} with gridless local search on the DML function, \subref{fig:varySNR_L=3_gridlessMAP} with gridless local search on the MAP function.}
	\label{fig:varySNR_L=3}
\end{figure}

\begin{figure}[t]
	\centering
	\ref{legend1}

	\subfigure[\label{fig:varyN_L=3_noGridless}]{\begin{tikzpicture}[inner ysep=1.5pt]
	\def\filepath{results/varyN_L=3_SNR=-5_uncorr_rmse.csv}
	
	\begin{axis}[%
		at={(0,0)},
		scale only axis,
		enlarge x limits={false},
		xlabel style={font=\color{white!15!black}, font=\scriptsize},
		xlabel={Number of snapshots $N$},
%		xlabel shift = -2.5pt,
		xmode = log,
		ylabel style={font=\color{white!15!black}, font=\scriptsize},
		ylabel={RMSE (rad)},
%		ylabel shift = -3pt,
		ymode = log,
		axis background/.style={fill=white},
		xmajorgrids,
		ymajorgrids,
		legend style={legend cell align=left, align=left, draw=white, % draw=white!15!black, 
			font=\scriptsize, row sep=-2pt,
			at={(0,0)}, anchor = south west},
		cycle multi list = {
%			solid,dashed,densely dotted\nextlist
%			[2 of]MATLABcolormarklist
			MATLABcolormarklist
%			MATLABcolorlist
		},
		legend to name=legend1,
		legend columns=3,
		]

		\addplot table [x index=0, y=SCIP-SDP] {\filepath};
		\addlegendentry{MISDP via SCIP-SDP};
		
%		\addplot[color=MATLABcolor6, densely dashdotted, mark=triangle, every mark/.append style = {solid}] table [x index=0, y=SCIP-SDP-SVD] {\filepath};
%		\addlegendentry{MISDP-S via SCIP-SDP};
%		
%		\pgfplotsset{cycle list shift=-1};
		
		\addplot table [x index=0, y=Pilanci] {\filepath};
		\addlegendentry{MISDP via RR};
		
%		\addplot[color=MATLABcolor7, densely dashdotted, mark=triangle, every mark/.append style = {solid}] table [x index=0, y=Pilanci-SVD] {\filepath};
%		\addlegendentry{MISDP-S via IRRR};
%		
%		\pgfplotsset{cycle list shift=-2};
		
		\addplot table [x index=0, y=SPARROW] {\filepath};
		\addlegendentry{SPARROW};
		
		\addplot table [x index=0, y=SBL] {\filepath};
		\addlegendentry{SBL};
		
		\addplot table [x index=0, y=MUSIC] {\filepath};
		\addlegendentry{MUSIC};

		\addplot+[color=TUDa-4a, densely dashdotted] table [x index=0, y=root-MUSIC] {\filepath};
		\addlegendentry{root-MUSIC};
		
		\addplot+[color=TUDa-0c, densely dashdotted,mark=null] table [x index=0, y=DML] {\filepath};
		\addlegendentry{DML};

		\addplot[color=black] table [x index=0, y=CRB] {\filepath};
		\addlegendentry{CRB};
		
%		\addplot+[dashed] table [x index=0, y=fmincon] {\filepath};
%		\addlegendentry{DML (\texttt{fmincon})};
%		\legend{};
		
	\end{axis}
\end{tikzpicture}} \\
	\subfigure[\label{fig:varyN_L=3_gridlessDML}]{\begin{tikzpicture}[inner ysep=1.5pt]
	\def\filepath{results/varyN_L=3_SNR=-5_uncorr_fminconDML_rmse.csv}
	
	\begin{axis}[%
		at={(0,0)},
		scale only axis,
		enlarge x limits={false},
		xlabel style={font=\color{white!15!black}, font=\scriptsize},
		xlabel={Number of snapshots $N$},
%		xlabel shift = -2.5pt,
		xmode = log,
		ylabel style={font=\color{white!15!black}, font=\scriptsize},
		ylabel={RMSE (rad)},
%		ylabel shift = -3pt,
		ymode = log,
		axis background/.style={fill=white},
		xmajorgrids,
		ymajorgrids,
		legend style={legend cell align=left, align=left, draw=white, % draw=white!15!black, 
			font=\tiny, row sep=-2pt,
			at={(0,0)}, anchor = south west},
		cycle multi list = {
%			solid,dashed,densely dotted\nextlist
%			[2 of]MATLABcolormarklist
			MATLABcolormarklist
%			MATLABcolorlist
		},
%		legend to name=legend1,
%		legend columns=8,
		]

		\addplot table [x index=0, y=SCIP-SDP] {\filepath};
		\addlegendentry{MISDP via SCIP-SDP};
		
%		\addplot[color=MATLABcolor6, densely dashdotted, mark=triangle, every mark/.append style = {solid}] table [x index=0, y=SCIP-SDP-SVD] {\filepath};
%		\addlegendentry{MISDP-S via SCIP-SDP};
%		
%		\pgfplotsset{cycle list shift=-1};
		
		\addplot table [x index=0, y=Pilanci] {\filepath};
		\addlegendentry{MISDP via IRRR};
		
%		\addplot[color=MATLABcolor7, densely dashdotted, mark=triangle, every mark/.append style = {solid}] table [x index=0, y=Pilanci-SVD] {\filepath};
%		\addlegendentry{MISDP-S via IRRR};
%		
%		\pgfplotsset{cycle list shift=-2};
		
		\addplot table [x index=0, y=SPARROW] {\filepath};
		\addlegendentry{SPARROW};
		
		\addplot table [x index=0, y=SBL] {\filepath};
		\addlegendentry{SBL};
		
		\addplot table [x index=0, y=MUSIC] {\filepath};
		\addlegendentry{MUSIC};
		
		\addplot+[color=TUDa-4a, densely dashdotted] table [x index=0, y=root-MUSIC] {\filepath};
		\addlegendentry{root-MUSIC};
		
		\addplot+[color=TUDa-0c, densely dashdotted,mark=null] table [x index=0, y=DML] {\filepath};
		\addlegendentry{DML};
		
		\addplot[color=black] table [x index=0, y=CRB] {\filepath};
		\addlegendentry{CRB};
		
%		\addplot+[dashed] table [x index=0, y=fmincon] {\filepath};
%		\addlegendentry{DML (\texttt{fmincon})};
		\legend{};
		
	\end{axis}
\end{tikzpicture}} \\
	\subfigure[\label{fig:varyN_L=3_gridlessMAP}]{\begin{tikzpicture}[inner ysep=1.5pt]
	\def\filepath{results/varyN_L=3_SNR=-5_uncorr_fminconMAP_rmse.csv}
	
	\begin{axis}[%
		at={(0,0)},
		scale only axis,
		enlarge x limits={false},
		xlabel style={font=\color{white!15!black}, font=\scriptsize},
		xlabel={Number of snapshots $N$},
%		xlabel shift = -2.5pt,
		xmode = log,
		ylabel style={font=\color{white!15!black}, font=\scriptsize},
		ylabel={RMSE (rad)},
%		ylabel shift = -3pt,
		ymode = log,
		axis background/.style={fill=white},
		xmajorgrids,
		ymajorgrids,
		legend style={legend cell align=left, align=left, draw=white, % draw=white!15!black, 
			font=\tiny, row sep=-2pt,
			at={(0,0)}, anchor = south west},
		cycle multi list = {
%			solid,dashed,densely dotted\nextlist
%			[2 of]MATLABcolormarklist
			MATLABcolormarklist
%			MATLABcolorlist
		},
%		legend to name=legend1,
%		legend columns=8,
		]

		\addplot table [x index=0, y=SCIP-SDP] {\filepath};
		\addlegendentry{MISDP via SCIP-SDP};
		
%		\addplot[color=MATLABcolor6, densely dashdotted, mark=triangle, every mark/.append style = {solid}] table [x index=0, y=SCIP-SDP-SVD] {\filepath};
%		\addlegendentry{MISDP-S via SCIP-SDP};
%		
%		\pgfplotsset{cycle list shift=-1};
		
		\addplot table [x index=0, y=Pilanci] {\filepath};
		\addlegendentry{MISDP via IRRR};
		
%		\addplot[color=MATLABcolor7, densely dashdotted, mark=triangle, every mark/.append style = {solid}] table [x index=0, y=Pilanci-SVD] {\filepath};
%		\addlegendentry{MISDP-S via IRRR};
%		
%		\pgfplotsset{cycle list shift=-2};
		
		\addplot table [x index=0, y=SPARROW] {\filepath};
		\addlegendentry{SPARROW};
		
		\addplot table [x index=0, y=SBL] {\filepath};
		\addlegendentry{SBL};
		
		\addplot table [x index=0, y=MUSIC] {\filepath};
		\addlegendentry{MUSIC};
		
		\addplot+[color=TUDa-4a, densely dashdotted] table [x index=0, y=root-MUSIC] {\filepath};
		\addlegendentry{root-MUSIC};
		
		\addplot+[color=TUDa-0c, densely dashdotted,mark=null] table [x index=0, y=DML] {\filepath};
		\addlegendentry{DML};
		
		\addplot[color=black] table [x index=0, y=CRB] {\filepath};
		\addlegendentry{CRB};
		
%		\addplot+[dashed] table [x index=0, y=fmincon] {\filepath};
%		\addlegendentry{DML (\texttt{fmincon})};
		\legend{};
		
	\end{axis}
\end{tikzpicture}}
	\caption{RMSE vs. the number of snapshots for $L=3$ uncorrelated sources, $M=8$ sensors, $\text{SNR} = -5$ dB, and $K=100$ grid points, \subref{fig:varyN_L=3_noGridless} without gridless local search, \subref{fig:varyN_L=3_gridlessDML} with gridless local search on the DML function, \subref{fig:varyN_L=3_gridlessMAP} with gridless local search on the MAP function.}
	\label{fig:varyN_L=3}
\end{figure}

%\begin{figure}[tb]
%	\centering
%	\input{images/Pilanci_continuousVar_L=3_SNR=10_uncorr.tikz}
%	\caption{Optimal fractional solutions $\widehat{\vect{u}}$ of the interval relaxation with different sample sizes for $L=5$ uncorrelated sources with spatial frequencies $\vect{\mu} = \pi \cdot [-0.5,-0.1, 0.35, 0.5,0.7]^\Trans$, $M=8$ sensors, $\text{SNR} = -5$ dB, and $K=100$ grid points.}
%	\label{fig:fractionalSolution}
%\end{figure}

%\begin{figure*}[t]
%	\centering
%	\input{images/varySNR_NGrd=1000_rmse.tikz}
%	\caption{RMSE vs SNR for $L=3$ uncorrelated sources, $M=8$ sensors, $N=8$ snapshots, and $K=1000$ grid points.}
%\end{figure*}

In Fig.~\ref{fig:varySNR_L=3} and~\ref{fig:varyN_L=3}, we compare the estimation error of the methods in a small-scale scenario of $L=3$ sources with spatial frequencies $\vect{\mu} = \pi \cdot [-0.1, 0.35, 0.47]^\Trans$, where the brute-force search is computationally competitive.
The true frequencies $\vect{\mu}$ are not forced to be on the searching grid.
The additional gridless local search based on the DML function and the MAP function in~\eqref{prob:MMVL20_original}, respectively, are performed to all the methods. Both the estimation errors before and after the gridless local search are reported.
% The estimation errors of the recovered frequencies without and with the additional gridless local search based on the DML function and the MAP function in~\eqref{prob:MMVL20_original}, respectively, are reported.

In \textit{Experiment 1} depicted in Fig.~\ref{fig:varySNR_L=3}, we fix $N=8$ and vary SNR from $-10$ to $20$ dB.
Except for root-MUSIC, all the methods are grid-based and, hence, as shown in Fig.~\ref{fig:varySNR_L=3_noGridless}, their estimation quality in the high SNR region is hindered by the grid mismatch error.
Nevertheless, in most cases, the grid mismatch error is eliminated by an additional gridless local search, as demonstrated in Fig.~\ref{fig:varySNR_L=3_gridlessDML} and~\ref{fig:varySNR_L=3_gridlessMAP}.
%Comparing Fig.~\ref{fig:varyN_L=3_noGridless} and~\ref{fig:varyN_L=3_gridlessDML} in the region of large sample size, we observe that the additional gridless local search eliminates the residual error caused by the finite grid as the RMSE of each method in Fig.~\ref{fig:varyN_L=3_gridlessDML} after the gridless local search asymptotically converges to the CRB.
% With the increase of SNR, the solution of SPARROW becomes less sparse than the expected value since the recommended choice of the regularization parameter $\lambda$ in~\eqref{eq:lambda} decreases.
The heuristic choice~\eqref{eq:lambda} for $\lambda$ in SPARROW is satisfactory for low SNR but fails to provide a sparse solution when the noise variance is small.
This leads to a significant degradation of the error performance of SPARROW for high SNR.
A more suitable choice of $\lambda$ may be obtained by cross validation with training data, which is not investigated in this paper.
Compared to SCIP-SDP, the RR algorithm finds a satisfactory approximate solution of the MISDP~\eqref{prob:MMVL20MISDP} or~\eqref{prob:MMVL20MISDP2}.
The proposed MISDP-based method and the brute-force DML exhibit similar SNR thresholds at which their RMSEs achieve the CRB, superior to those of the other methods.
% Moreover, the proposed MISDP-based method exhibits a lower SNR threshold, at which the RMSE achieves the CRB, than that of the other methods.
% Moreover, MISDP has the best estimation quality in the region before the threshold.
%Compared to the brute-force DML, the SBL method exhibits similar estimation quality in the regions before and after the threshold, whereas its threshold occurs at a higher SNR, i.e., the RMSE of the SBL method deviates from the CRB at a higher SNR.

In \textit{Experiment 2} shown in Fig.~\ref{fig:varyN_L=3}, we fix SNR $= -5$ dB and vary the number of snapshots $ N $ from $1$ to $10^3$.
%As discussed in Section~\ref{sec:model}, to achieve the lower running time in the SDP implementation, the MISDP reformulation~\eqref{prob:MMVL20MISDP} is employed in the undersampled case, whereas the equivalent MISDP reformulation~\eqref{prob:MMVL20MISDP2}, whose constraint dimension is independent of the number of snapshots, is chosen in the oversampled case.
Similar to Fig.~\ref{fig:varySNR_L=3}, Fig.~\ref{fig:varyN_L=3} demonstrates that the RR algorithm finds a satisfactory approximate solution of the MISDP problem~\eqref{prob:MMVL20MISDP} or~\eqref{prob:MMVL20MISDP2} in the inspected sample size region, as compared to the exact solution obtained by SCIP-SDP.

The brute-force DML and the proposed method exhibit similar threshold performance, which is superior to that of the other methods.
As stated in Section~\ref{sec:bayesian}, the MAP estimation problem differs from the DML only in the additional Tikhonov regularization introduced by the Gaussian prior.
Compared to the unbiased DML estimator, the use of additional knowledge of the waveform's prior distribution in the MAP estimator typically leads to a decreased estimation variance but an increased bias, which is rooted in the well-known bias-variance trade-off~\cite[Sec. 6.2]{jamesIntroductionStatisticalLearning2021}.
%The grid mismatch error is removed by the gridless local search for the DML and MAP estimators in Fig.~\ref{fig:varyN_L=3_gridlessDML} and~\ref{fig:varyN_L=3_gridlessMAP}, respectively.
As revealed in Fig.~\ref{fig:varyN_L=3_noGridless}, the proposed MAP-based method has a lower estimation error than DML in the region of low sample sizes but, for large sample sizes, the RMSE of the proposed method is dominated by the asymptotic bias.
%as the grid mismatch error is removed by the gridless local search.
% Nevertheless, as shown in Fig.~\ref{fig:varyN_L=3_gridlessDML}, the proposed method still provides a good initialization for the gridless local search on the DML function in the asymptotic region.
Nevertheless, for most of the methods, both the bias introduced by the regularization and the residual error caused by the finite grid in the asymptotic region can be eliminated by the additional gridless local search on the DML function.
%In comparison to SCIP-SDP, the IRRR algorithm is shown in Fig.~\ref{fig:varySNR_L=3} to be capable of obtaining a good approximate solution of the proposed MISDP problem in~\eqref{prob:MMVL20MISDP} in the region of low sample sizes, whereas, for large sample sizes in Fig.~\ref{fig:varyN_L=3}, it exhibits a significant degradation of the estimation quality since the interval relaxation becomes less tight, which is illustrated in Fig.~\ref{fig:fractionalSolution}.
%Nevertheless, Fig.~\ref{fig:varyN_L=3_gridlessDML} demonstrates that the proposed method via IRRR can be used to find a good initialization for the gridless local search in the asymptotic region. In particular, with the additional gridless local search, the proposed method via IRRR achieves the CRB at the same value of $N$ as root-MUSIC, which is superior to SPARROW and MUSIC but inferior to the brute-force DML and the proposed method via SCIP-SDP, and, moreover, the proposed method via IRRR obtains a lower estimation error than that of root-MUSIC before the threshold.

% Different from the results in Fig.~\ref{fig:varySNR_L=3}, the threshold of SBL occurs at a higher number of snapshots than that of DML.
Due to the continuous relaxation and the Lagrangian relaxation discussed in Section~\ref{sec:SPARROW}, SPARROW is outperformed by DML and the proposed MISDP method. In particular, it is observed to have an inferior resolution in the next experiment in Fig.~\ref{fig:varyFreq_L=3} and, hence, it fails to recover the three closely located sources in the inspected sample size region in Fig.~\ref{fig:varyN_L=3}.

\begin{figure}[t]
	\centering
	\ref{legend1}

	\subfigure[\label{fig:varyFreq_L=3_noGridless}]{
		\begin{tikzpicture}[inner ysep=1.5pt]
  \def\filepath{results/varyFreq_L=3_M=8_N=8_SNR=10_uncorr_rmse.csv}
	
	\begin{axis}[%
		at={(0,0)},
		scale only axis,
		enlarge x limits={false},
		xlabel style={font=\color{white!15!black}, font=\scriptsize},
		xlabel={Frequency separation $ \delta \pi $ (rad)},
%		xlabel shift = -2.5pt,
%		xmode = log,
%		xtick = {0.025,0.05,0.1,0.2,0.4},
%		xticklabels = {0.025,0.05,0.1,0.2,0.4},
    % scaled x ticks = {real:3.1415926},
    % xtick scale label code/.code={$\cdot \pi$},
		%ymin=0,
%		ymax=1,
		ylabel style={font=\color{white!15!black}, font=\scriptsize},
    ylabel={RMSE (rad)},
%		ylabel shift = -3pt,
		ymode = log,
		axis background/.style={fill=white},
		xmajorgrids,
		ymajorgrids,
		legend style={legend cell align=left, align=left, draw=white, % draw=white!15!black, 
			font=\tiny, row sep=-2pt,
			at={(0,0)}, anchor = south west},
		cycle multi list = {
%			solid,dashed,densely dotted\nextlist
%			[2 of]MATLABcolormarklist
			MATLABcolormarklist
%			MATLABcolorlist
		},
	% legend to name=legend1,
	% legend columns=8,
		]

		\addplot table [x index=0, y=SCIP-SDP] {\filepath};
		\addlegendentry{MISDP via SCIP-SDP};
		
%		\addplot[color=MATLABcolor6, densely dashdotted, mark=triangle, every mark/.append style = {solid}] table [x index=0, y=SCIP-SDP-SVD] {\filepath};
%		\addlegendentry{MISDP-S via SCIP-SDP};
%		
%		\pgfplotsset{cycle list shift=-1};
		
		\addplot table [x index=0, y=Pilanci] {\filepath};
		\addlegendentry{MISDP via RR};
		
%		\addplot[color=MATLABcolor7, densely dashdotted, mark=triangle, every mark/.append style = {solid}] table [x index=0, y=Pilanci-SVD] {\filepath};
%		\addlegendentry{MISDP-S via IRRR};
%		
%		\pgfplotsset{cycle list shift=-2};
		
		\addplot table [x index=0, y=SPARROW] {\filepath};
		\addlegendentry{SPARROW};
		
		\addplot table [x index=0, y=SBL] {\filepath};
		\addlegendentry{SBL};
		
		\addplot table [x index=0, y=MUSIC] {\filepath};
		\addlegendentry{MUSIC};
		
		\addplot+[color=TUDa-4a, densely dashdotted] table [x index=0, y=root-MUSIC] {\filepath};
		\addlegendentry{root-MUSIC};
		
		\addplot+[color=TUDa-0c, densely dashdotted,mark=null] table [x index=0, y=DML] {\filepath};
		\addlegendentry{DML};
		
		\addplot[color=black] table [x index=0, y=CRB] {\filepath};
		\addlegendentry{CRB};
		
%		\addplot+[dashed] table [x index=0, y=fmincon] {\filepath};
%		\addlegendentry{DML (\texttt{fmincon})};
		\legend{};
		
	\end{axis}
\end{tikzpicture}
	} \\
	\subfigure[\label{fig:varyFreq_L=3_gridlessDML}]{\begin{tikzpicture}[inner ysep=1.5pt]
  \def\filepath{results/varyFreq_L=3_M=8_N=8_SNR=10_uncorr_fminconDML_rmse.csv}
	
	\begin{axis}[%
		at={(0,0)},
		scale only axis,
		enlarge x limits={false},
		xlabel style={font=\color{white!15!black}, font=\scriptsize},
    xlabel={Frequency separation $ \delta \pi $ (rad)},
%		xlabel shift = -2.5pt,
%		xmode = log,
%		xtick = {0.025,0.05,0.1,0.2,0.4},
%		xticklabels = {0.025,0.05,0.1,0.2,0.4},
		%ymin=0,
%		ymax=1,
		ylabel style={font=\color{white!15!black}, font=\scriptsize},
		ylabel={RMSE (rad)},
%		ylabel shift = -3pt,
		ymode = log,
		axis background/.style={fill=white},
		xmajorgrids,
		ymajorgrids,
		legend style={legend cell align=left, align=left, draw=white, % draw=white!15!black, 
			font=\tiny, row sep=-2pt,
			at={(0,0)}, anchor = south west},
		cycle multi list = {
%			solid,dashed,densely dotted\nextlist
%			[2 of]MATLABcolormarklist
			MATLABcolormarklist
%			MATLABcolorlist
		},
%		legend to name=legend1,
%		legend columns=8,
		]

		\addplot table [x index=0, y=SCIP-SDP] {\filepath};
		\addlegendentry{MISDP via SCIP-SDP};
		
%		\addplot[color=MATLABcolor6, densely dashdotted, mark=triangle, every mark/.append style = {solid}] table [x index=0, y=SCIP-SDP-SVD] {\filepath};
%		\addlegendentry{MISDP-S via SCIP-SDP};
%		
%		\pgfplotsset{cycle list shift=-1};
		
		\addplot table [x index=0, y=Pilanci] {\filepath};
		\addlegendentry{MISDP via IRRR};
		
%		\addplot[color=MATLABcolor7, densely dashdotted, mark=triangle, every mark/.append style = {solid}] table [x index=0, y=Pilanci-SVD] {\filepath};
%		\addlegendentry{MISDP-S via IRRR};
%		
%		\pgfplotsset{cycle list shift=-2};
		
		\addplot table [x index=0, y=SPARROW] {\filepath};
		\addlegendentry{SPARROW};
		
		\addplot table [x index=0, y=SBL] {\filepath};
		\addlegendentry{SBL};
		
		\addplot table [x index=0, y=MUSIC] {\filepath};
		\addlegendentry{MUSIC};
		
		\addplot+[color=TUDa-4a, densely dashdotted] table [x index=0, y=root-MUSIC] {\filepath};
		\addlegendentry{root-MUSIC};
		
		\addplot+[color=TUDa-0c, densely dashdotted,mark=null] table [x index=0, y=DML] {\filepath};
		\addlegendentry{DML};
		
		\addplot[color=black] table [x index=0, y=CRB] {\filepath};
		\addlegendentry{CRB};
		
%		\addplot+[dashed] table [x index=0, y=fmincon] {\filepath};
%		\addlegendentry{DML (\texttt{fmincon})};
		\legend{};
		
	\end{axis}
\end{tikzpicture}} \\
	\subfigure[\label{fig:varyFreq_L=3_gridlessMAP}]{\begin{tikzpicture}[inner ysep=1.5pt]
  \def\filepath{results/varyFreq_L=3_M=8_N=8_SNR=10_uncorr_fminconMAP_rmse.csv}
	
	\begin{axis}[%
		at={(0,0)},
		scale only axis,
		enlarge x limits={false},
		xlabel style={font=\color{white!15!black}, font=\scriptsize},
    xlabel={Frequency separation $ \delta \pi $ (rad)},
%		xlabel shift = -2.5pt,
%		xmode = log,
%		xtick = {0.025,0.05,0.1,0.2,0.4},
%		xticklabels = {0.025,0.05,0.1,0.2,0.4},
		%ymin=0,
%		ymax=1,
		ylabel style={font=\color{white!15!black}, font=\scriptsize},
		ylabel={RMSE (rad)},
%		ylabel shift = -3pt,
		ymode = log,
		axis background/.style={fill=white},
		xmajorgrids,
		ymajorgrids,
		legend style={legend cell align=left, align=left, draw=white, % draw=white!15!black, 
			font=\tiny, row sep=-2pt,
			at={(0,0)}, anchor = south west},
		cycle multi list = {
%			solid,dashed,densely dotted\nextlist
%			[2 of]MATLABcolormarklist
			MATLABcolormarklist
%			MATLABcolorlist
		},
%		legend to name=legend1,
%		legend columns=8,
		]

		\addplot table [x index=0, y=SCIP-SDP] {\filepath};
		\addlegendentry{MISDP via SCIP-SDP};
		
%		\addplot[color=MATLABcolor6, densely dashdotted, mark=triangle, every mark/.append style = {solid}] table [x index=0, y=SCIP-SDP-SVD] {\filepath};
%		\addlegendentry{MISDP-S via SCIP-SDP};
%		
%		\pgfplotsset{cycle list shift=-1};
		
		\addplot table [x index=0, y=Pilanci] {\filepath};
		\addlegendentry{MISDP via IRRR};
		
%		\addplot[color=MATLABcolor7, densely dashdotted, mark=triangle, every mark/.append style = {solid}] table [x index=0, y=Pilanci-SVD] {\filepath};
%		\addlegendentry{MISDP-S via IRRR};
%		
%		\pgfplotsset{cycle list shift=-2};
		
		\addplot table [x index=0, y=SPARROW] {\filepath};
		\addlegendentry{SPARROW};
		
		\addplot table [x index=0, y=SBL] {\filepath};
		\addlegendentry{SBL};
		
		\addplot table [x index=0, y=MUSIC] {\filepath};
		\addlegendentry{MUSIC};
		
		\addplot+[color=TUDa-4a, densely dashdotted] table [x index=0, y=root-MUSIC] {\filepath};
		\addlegendentry{root-MUSIC};
		
		\addplot+[color=TUDa-0c, densely dashdotted,mark=null] table [x index=0, y=DML] {\filepath};
		\addlegendentry{DML};
		
		\addplot[color=black] table [x index=0, y=CRB] {\filepath};
		\addlegendentry{CRB};
		
%		\addplot+[dashed] table [x index=0, y=fmincon] {\filepath};
%		\addlegendentry{DML (\texttt{fmincon})};
		\legend{};
		
	\end{axis}
\end{tikzpicture}}
	\caption{RMSE vs. frequency separation for $L=3$ uncorrelated sources, two of which are separated by $ \delta $, $M=8$ sensors, $N=8$ snapshots, SNR $ = 10 $ dB and $K=100$ grid points, \subref{fig:varyFreq_L=3_noGridless} without gridless local search, \subref{fig:varyFreq_L=3_gridlessDML} with gridless local search on the DML function, \subref{fig:varyFreq_L=3_gridlessMAP} with gridless local search on the MAP function.}
	\label{fig:varyFreq_L=3}
\end{figure}

In \textit{Experiment 3}, in Fig.~\ref{fig:varyFreq_L=3}, we investigate the resolution performance of the methods by varying the frequency separation between two of the $L=3$ sources. The frequencies are given by $\vect{\mu} = \pi \cdot [-0.1, 0.35, 0.35 + \delta]^\Trans$ with $\delta$ varied from $0.01$ to $0.15$. We fix SNR $=10$ dB and $N=8$. MISDP achieves the CRB for very closely located sources and, hence, it has the best resolution among all the methods. Again, the RR algorithm provides a good approximate solution of the MISDP problem even for closely located sources.

\begin{figure}[t]
	\centering
	\ref{legend1}

	\subfigure[]{\begin{tikzpicture}[inner ysep=1.5pt]
	\def\filepath{results/varyN_L=5_SNR=-5_uncorr_rmse.csv}
	
	\begin{axis}[%
		at={(0,0)},
		scale only axis,
		enlarge x limits={false},
		xlabel style={font=\color{white!15!black}, font=\scriptsize},
		xlabel={Number of snapshots $N$},
%		xlabel shift = -2.5pt,
		xmode = log,
		ylabel style={font=\color{white!15!black}, font=\scriptsize},
		ylabel={RMSE (rad)},
%		ylabel shift = -3pt,
		ymode = log,
		axis background/.style={fill=white},
		xmajorgrids,
		ymajorgrids,
		legend style={legend cell align=left, align=left, draw=white, % draw=white!15!black, 
			font=\tiny, row sep=-2pt,
			at={(0,0)}, anchor = south west},
		cycle multi list = {
%			solid,dashed,densely dotted\nextlist
%			[2 of]MATLABcolormarklist
			MATLABcolormarklist
%			MATLABcolorlist
		},
%		legend to name=legend1,
%		legend columns=8,
		]

		\addplot table [x index=0, y=SCIP-SDP] {\filepath};
		\addlegendentry{MISDP via SCIP-SDP};
		
%		\addplot[color=MATLABcolor6, densely dashdotted, mark=triangle, every mark/.append style = {solid}] table [x index=0, y=SCIP-SDP-SVD] {\filepath};
%		\addlegendentry{MISDP-S via SCIP-SDP};
%		
%		\pgfplotsset{cycle list shift=-1};
		
		\addplot table [x index=0, y=Pilanci] {\filepath};
		\addlegendentry{MISDP via IRRR};
		
%		\addplot[color=MATLABcolor7, densely dashdotted, mark=triangle, every mark/.append style = {solid}] table [x index=0, y=Pilanci-SVD] {\filepath};
%		\addlegendentry{MISDP-S via IRRR};
%		
%		\pgfplotsset{cycle list shift=-2};
		
		\addplot table [x index=0, y=SPARROW] {\filepath};
		\addlegendentry{SPARROW};
		
		\addplot table [x index=0, y=SBL] {\filepath};
		\addlegendentry{SBL};
		
		\addplot table [x index=0, y=MUSIC] {\filepath};
		\addlegendentry{MUSIC};
		
		\addplot+[color=TUDa-4a, densely dashdotted] table [x index=0, y=root-MUSIC] {\filepath};
		\addlegendentry{root-MUSIC};
		
		\addplot+[color=TUDa-0c, densely dashdotted,mark=null] table [x index=0, y=DML] {\filepath};
		\addlegendentry{DML};
		
		\addplot[color=black] table [x index=0, y=CRB] {\filepath};
		\addlegendentry{CRB};
		
%		\addplot+[dashed] table [x index=0, y=fmincon] {\filepath};
%		\addlegendentry{DML (\texttt{fmincon})};
		\legend{};
		
	\end{axis}
\end{tikzpicture}} \\
	\subfigure[]{\begin{tikzpicture}[inner ysep=1.5pt]
	\def\filepath{results/varyN_L=5_SNR=-5_uncorr_fminconDML_rmse.csv}
	
	\begin{axis}[%
		at={(0,0)},
		scale only axis,
		enlarge x limits={false},
		xlabel style={font=\color{white!15!black}, font=\scriptsize},
		xlabel={Number of snapshots $N$},
%		xlabel shift = -2.5pt,
		xmode = log,
		ylabel style={font=\color{white!15!black}, font=\scriptsize},
		ylabel={RMSE (rad)},
%		ylabel shift = -3pt,
		ymode = log,
		axis background/.style={fill=white},
		xmajorgrids,
		ymajorgrids,
		legend style={legend cell align=left, align=left, draw=white, % draw=white!15!black, 
			font=\tiny, row sep=-2pt,
			at={(0,0)}, anchor = south west},
		cycle multi list = {
%			solid,dashed,densely dotted\nextlist
%			[2 of]MATLABcolormarklist
			MATLABcolormarklist
%			MATLABcolorlist
		},
%		legend to name=legend1,
%		legend columns=8,
		]

		\addplot table [x index=0, y=SCIP-SDP] {\filepath};
		\addlegendentry{MISDP via SCIP-SDP};
		
%		\addplot[color=MATLABcolor6, densely dashdotted, mark=triangle, every mark/.append style = {solid}] table [x index=0, y=SCIP-SDP-SVD] {\filepath};
%		\addlegendentry{MISDP-S via SCIP-SDP};
%		
%		\pgfplotsset{cycle list shift=-1};
		
		\addplot table [x index=0, y=Pilanci] {\filepath};
		\addlegendentry{MISDP via IRRR};
		
%		\addplot[color=MATLABcolor7, densely dashdotted, mark=triangle, every mark/.append style = {solid}] table [x index=0, y=Pilanci-SVD] {\filepath};
%		\addlegendentry{MISDP-S via IRRR};
%		
%		\pgfplotsset{cycle list shift=-2};
		
		\addplot table [x index=0, y=SPARROW] {\filepath};
		\addlegendentry{SPARROW};
		
		\addplot table [x index=0, y=SBL] {\filepath};
		\addlegendentry{SBL};
		
		\addplot table [x index=0, y=MUSIC] {\filepath};
		\addlegendentry{MUSIC};
		
		\addplot+[color=TUDa-4a, densely dashdotted] table [x index=0, y=root-MUSIC] {\filepath};
		\addlegendentry{root-MUSIC};
		
		\addplot+[color=TUDa-0c, densely dashdotted,mark=null] table [x index=0, y=DML] {\filepath};
		\addlegendentry{DML};
		
		\addplot[color=black] table [x index=0, y=CRB] {\filepath};
		\addlegendentry{CRB};
		
%		\addplot+[dashed] table [x index=0, y=fmincon] {\filepath};
%		\addlegendentry{DML (\texttt{fmincon})};
		\legend{};
		
	\end{axis}
\end{tikzpicture}} \\
	\subfigure[\label{fig:varyN_L=5_gridlessMAP}]{\begin{tikzpicture}[inner ysep=1.5pt]
	\def\filepath{results/varyN_L=5_SNR=-5_uncorr_fminconMAP_rmse.csv}
	
	\begin{axis}[%
		at={(0,0)},
		scale only axis,
		enlarge x limits={false},
		xlabel style={font=\color{white!15!black}, font=\scriptsize},
		xlabel={Number of snapshots $N$},
%		xlabel shift = -2.5pt,
		xmode = log,
		ylabel style={font=\color{white!15!black}, font=\scriptsize},
		ylabel={RMSE (rad)},
%		ylabel shift = -3pt,
		ymode = log,
		axis background/.style={fill=white},
		xmajorgrids,
		ymajorgrids,
		legend style={legend cell align=left, align=left, draw=white, % draw=white!15!black, 
			font=\tiny, row sep=-2pt,
			at={(0,0)}, anchor = south west},
		cycle multi list = {
%			solid,dashed,densely dotted\nextlist
%			[2 of]MATLABcolormarklist
			MATLABcolormarklist
%			MATLABcolorlist
		},
%		legend to name=legend1,
%		legend columns=8,
		]

		\addplot table [x index=0, y=SCIP-SDP] {\filepath};
		\addlegendentry{MISDP via SCIP-SDP};
		
%		\addplot[color=MATLABcolor6, densely dashdotted, mark=triangle, every mark/.append style = {solid}] table [x index=0, y=SCIP-SDP-SVD] {\filepath};
%		\addlegendentry{MISDP-S via SCIP-SDP};
%		
%		\pgfplotsset{cycle list shift=-1};
		
		\addplot table [x index=0, y=Pilanci] {\filepath};
		\addlegendentry{MISDP via IRRR};
		
%		\addplot[color=MATLABcolor7, densely dashdotted, mark=triangle, every mark/.append style = {solid}] table [x index=0, y=Pilanci-SVD] {\filepath};
%		\addlegendentry{MISDP-S via IRRR};
%		
%		\pgfplotsset{cycle list shift=-2};
		
		\addplot table [x index=0, y=SPARROW] {\filepath};
		\addlegendentry{SPARROW};
		
		\addplot table [x index=0, y=SBL] {\filepath};
		\addlegendentry{SBL};
		
		\addplot table [x index=0, y=MUSIC] {\filepath};
		\addlegendentry{MUSIC};
		
		\addplot+[color=TUDa-4a, densely dashdotted] table [x index=0, y=root-MUSIC] {\filepath};
		\addlegendentry{root-MUSIC};
		
		\addplot+[color=TUDa-0c, densely dashdotted,mark=null] table [x index=0, y=DML] {\filepath};
		\addlegendentry{DML};
		
		\addplot[color=black] table [x index=0, y=CRB] {\filepath};
		\addlegendentry{CRB};
		
%		\addplot+[dashed] table [x index=0, y=fmincon] {\filepath};
%		\addlegendentry{DML (\texttt{fmincon})};
		\legend{};
		
	\end{axis}
\end{tikzpicture}}
	\caption{RMSE vs. the number of snapshots for $L=5$ uncorrelated sources, $M=8$ sensors, $\text{SNR} = -5$ dB, and $K=100$ grid points, (a) without gridless local search, (b) with gridless local search on the DML function, (c) with gridless local search on the MAP function.}
	\label{fig:varyN_L=5}
\end{figure}

%\begin{table}[t]
%	\caption{Average computation time in seconds of different methods}
%	\label{tab:time}
%	\centering
%	\begin{tabular}{|c|c|c|c|c|}
%		\hline
%		DML & SCIP-SDP & Pilanci & MUSIC & root-MUSIC \\
%		\hline
%		490 & 20 &  & & \\
%		\hline
%	\end{tabular}
%\end{table}

\begin{figure}[t]
	\centering
	\ref{legend2}

	\begin{tikzpicture}[inner ysep=1.5pt]
	\def\filepath{results/varyN_L=5_SNR=-5_uncorr_runtime}
	
	\begin{axis}[%
%		width=0.45\columnwidth,
		%height=2cm,
		at={(0,0)},
		scale only axis,
		xmin=1,
		xmax=1000,
		xlabel style={font=\color{white!15!black}, font=\scriptsize},
		xlabel={Number of snapshots $N$},
%		xlabel shift = -2.5pt,
		xmode = log,
		ylabel style={font=\color{white!15!black}, font=\scriptsize},
		ylabel={Computation time (s)},
%		ylabel shift = -3pt,
		ymode = log,
		axis background/.style={fill=white},
		xmajorgrids,
		ymajorgrids,
		legend style={legend cell align=left, align=left, draw=none,
			% draw=white!15!black, 
			font=\scriptsize, row sep=-2pt,
			legend pos=outer north east,
%			at={(0,0)}, anchor = south west,
		},
		cycle multi list = {
%			solid,dashed,densely dotted\nextlist
%			[2 of]MATLABcolormarklist
			MATLABcolormarklist
%			MATLABcolorlist
		},
		legend to name=legend2,
		legend columns=2,
		]

		\addplot table [x index=0, y=SCIP-SDP-N] {\filepath_2.csv};
		\addlegendentry{MISDP~\eqref{prob:MMVL20MISDP} via SCIP-SDP};
		\addplot[color=MATLABcolor6, densely dashdotted, mark=triangle, every mark/.append style = {solid}] table [x index=0, y=SCIP-SDP-M] {\filepath.csv};
		\addlegendentry{MISDP~\eqref{prob:MMVL20MISDP2} via SCIP-SDP};
		
		\pgfplotsset{cycle list shift=-1};
		
		\addplot table [x index=0, y=Pilanci-N] {\filepath_2.csv};
		\addlegendentry{MISDP~\eqref{prob:MMVL20MISDP} via RR};
		\addplot[color=MATLABcolor7, densely dashdotted, mark=triangle, every mark/.append style = {solid}] table [x index=0, y=Pilanci-M] {\filepath.csv};
		\addlegendentry{MISDP~\eqref{prob:MMVL20MISDP2} via RR};
		
		\pgfplotsset{cycle list shift=-2};
		
		\addplot table [x index=0, y=SPARROW] {\filepath.csv};
		\addlegendentry{SPARROW};
		
		\addplot table [x index=0, y=SBL] {\filepath.csv};
		\addlegendentry{SBL};
%		\pgfplotsset{cycle list shift=0};
		
		\addplot table [x index=0, y=MUSIC] {\filepath.csv};
		\addlegendentry{MUSIC};

		\addplot+[color=TUDa-4a, densely dashdotted] table [x index=0, y=root-MUSIC] {\filepath.csv};
		\addlegendentry{root-MUSIC};
		
		\addplot[color=TUDa-0c, densely dashdotted] table [x index=0, y=DML] {\filepath.csv};
		\addlegendentry{DML};
		
%		\addplot+[dashed] table [x index=0, y=fmincon] {\filepath};
%		\addlegendentry{DML (\texttt{fmincon})};
%		\legend{};
		
	\end{axis}
\end{tikzpicture}
	\caption{Computation time vs. the number of snapshots for $L=5$ uncorrelated sources, $M=8$ sensors, $\text{SNR} = -5$ dB, and $K=100$ grid points.}
	\label{fig:varyN_L=5_runtime}
\end{figure}

In \textit{Experiment 4}, we consider $L=5$ sources with frequencies $\vect{\mu} = \pi \cdot [-0.5, 0.1, 0.35, 0.5, 0.7]^\Trans$, which is more difficult as the size of the binary feasible set of problem~\eqref{prob:MMVL20MISDP} increases exponentially with $L$. Hence, we increase the number $T$ of randomly generated binary solutions to $10^5$ for the RR Algorithm~\ref{algo1}.
Moreover, the solution obtained by Algorithm~\ref{algo1} is chosen to be the initialization for SCIP-SDP, which provides a better initial upper bound of the optimal value and, consequently, accelerates the branch-and-bound process in SCIP-SDP.
% \footnote{One may alternatively increase the number of randomly generated binary solutions in the internal randomized rounding procedure in SCIP-SDP, which is, however, impossible with the current version of the MATLAB interface of the SCIP-SDP solver.}
%Limited by the MATLAB interface of the SCIP-SDP solver, we cannot change the parameters of the internal randomized rounding procedure in SCIP-SDP. Thus, the solution obtained by Algorithm~\ref{algo1} is chosen to be the initialization for SCIP-SDP, which, however, is similar to setting $T=10^5$ for the randomized rounding procedure at the root of the branch-and-bound tree in SCIP-SDP.
%In this case, neither the brute-force DML nor SCIP-SDP converges in a reasonable time.
The error performance is reported in Fig.~\ref{fig:varyN_L=5} and the computation time in Fig.~\ref{fig:varyN_L=5_runtime}.
The complexities of the equivalent MISDPs~\eqref{prob:MMVL20MISDP} and~\eqref{prob:MMVL20MISDP2} are compared when solved by the SCIP-SDP solver and the RR Algorithm~\ref{algo1}.
Due to the high computation time of the brute-force DML, the results are averaged over only $N_R = 200$ Monte-Carlo trials.
The computation time of the additional gridless local search by \texttt{fmincon} is negligible compared to SPARROW and, hence, is not reported.
Although branch-and-bound enjoys improved scalability compared to brute-force search,
to limit the total execution time, in this experiment and Experiment 5, we terminate SCIP-SDP when 500 branch-and-bound nodes are explored or a time limit of 600 seconds is achieved.
%Note that, compared to the MISDP problem~\eqref{prob:MMVL20MISDP}, the change of the dimension of the semidefinite constraint in the equivalent reformulation~\eqref{prob:MMVL20MISDP2} only affects the running time of the relaxed continuous SDP subproblems in the SCIP-SDP solver, not the searching path over the branch-and-bound tree.
%Therefore, the difference in the complexities of the two equivalent MISDP reformulations~\eqref{prob:MMVL20MISDP} and~\eqref{prob:MMVL20MISDP2} can be analytically observed by comparing the dimensions of the semidefinite constraints.
%Thus, in this simulation, we do not compare the computation time among the two equivalent MISDP reformulations~\eqref{prob:MMVL20MISDP} and~\eqref{prob:MMVL20MISDP2}.
%Instead, as discussed in Section~\ref{sec:model}, to achieve the lower running time in the SDP implementation, the MISDP reformulation~\eqref{prob:MMVL20MISDP} is employed in the undersampled case, whereas the equivalent MISDP reformulation~\eqref{prob:MMVL20MISDP2}, whose constraint dimension is independent of the number of snapshots, is chosen in the oversampled case.
%The result is that, in Fig.~\ref{fig:varyN_L=5_runtime}, the computation time of the proposed MISDP-based method via SCIP-SDP increases with the increase of the number of snapshots in the undersampled case but remains constant in the oversampled case.
%Nevertheless, compared to SCIP-SDP, which uses 10 minutes, Pilanci's method and SPARROW exhibit a significantly reduced complexity, which require on average less than $2$ seconds.

Compared to Fig.~\ref{fig:varyN_L=3}, two major differences are observed in Fig.~\ref{fig:varyN_L=5}.
First, as mentioned in Section~\ref{sec:solution}, due to the incorporation of a randomized rounding procedure in the branch-and-bound approach, the SCIP-SDP solver often quickly finds a nearly optimal solution but spends much more time improving the lower bounds to verify its optimality.
Therefore, although only 500 branch-and-bound nodes are visited, MISDP via SCIP-SDP presents a more significant decrease of the RMSE compared to DML in the region of a low sample size than that in Fig.~\ref{fig:varyN_L=3}, which leads to a threshold performance superior to DML.
In fact, even without the branch-and-bound, the RR Algorithm~\ref{algo1} already finds a solution of similar estimation quality for a proper choice of $T$.
This suggests that, for a low sample size, MISDP via RR is more favorable than the brute-force DML since it possesses not only a superior error performance but also a reduced running time, as shown in Fig.~\ref{fig:varyN_L=5_runtime}.
Second, all the other methods, except for DML and MISDP, exhibit a degradation of error performance.
%In particular, the proposed method via IRRR fails in the region of large sample size, even with an additional gridless local search. Actually, when applied to the solutions of the IRRR algorithm, the gridless local search worsens the estimation quality in the asymptotic region as the frequencies recovered by the IRRR algorithm become meaningless.
%However, in the case of very few snapshots, the proposed method via IRRR is competitive due to its low complexity and satisfactory error performance.
Moreover, the MAP estimation model incorporates the additional prior information of the source covariance matrix, which is assumed to be unknown in the stochastic model for the CRB. Consequently, the proposed method via SCIP-SDP achieves an RMSE below the CRB in Fig.~\ref{fig:varyN_L=5_gridlessMAP} in the middle region.
%Finally, we remark that the SCIP-SDP solver incorporates a strategy, similar to that in the IRRR algorithm, that searches for good integer solutions by randomized rounding after a relaxed SDP subproblem is solved. This partly explains the observation that the SCIP-SDP exhibits a good error performance even with only 500 branch-and-bound nodes explored.

We then analyze the computation time of the two equivalent MISDP reformulations~\eqref{prob:MMVL20MISDP} and~\eqref{prob:MMVL20MISDP2} for the joint sparse MAP estimation. The problems~\eqref{prob:MMVL20MISDP} and~\eqref{prob:MMVL20MISDP2} differ only in the semidefinite constraint, in particular, in their dimensions. The computation time of an interior-point solver for continuous SDP problems generally increases dramatically with the increase of the dimensions of the semidefinite constraints. In particular, for both the RR Algorithm~\ref{algo1} and the SCIP-SDP solver, the difference in the dimension of the semidefinite constraints between the formulations~\eqref{prob:MMVL20MISDP} and~\eqref{prob:MMVL20MISDP2} only affects the computation time of the relaxed continuous SDP problem or subproblems in the branch-and-bound process. The randomized rounding step in Algorithm~\ref{algo1} and the searching path over the branch-and-bound tree are independent of the choice of the MISDP formulation.
In contrast to the semidefinite constraint~\eqref{eq:MMVL20MISDP_constraint}, the dimension of the constraint~\eqref{eq:MMVL20MISDP2_constraint} becomes independent of the number of snapshots $N$ by introducing an equivalent reformulated data matrix with $N=M$.

Consistent with the above analysis, in Fig.~\ref{fig:varyN_L=5_runtime}, the computation time of SCIP-SDP and the RR algorithm for the MISDP formulation~\eqref{prob:MMVL20MISDP} increases significantly with $ N $, whereas the computation time remains constant when the formulation~\eqref{prob:MMVL20MISDP2} is used.
However, considering the computation time of the randomized rounding step in Algorithm~\ref{algo1}, there is no significant difference in the total computation time of Algorithm~\ref{algo1} between the two formulations in the undersampled region.
Note that, a time limit of 600 seconds is set for the SCIP-SDP solver and the formulation~\eqref{prob:MMVL20MISDP} is only considered for a sample size up to $100$ due to high computation time.

\begin{figure}[t]
	\centering
	\ref{legend2}

	\begin{tikzpicture}[inner ysep=1.5pt]
  \def\filepath{results/varyM_N=50_L=5_SNR=-5_uncorr_runtime}
	
	\begin{axis}[%
%		width=0.45\columnwidth,
		%height=2cm,
		at={(0,0)},
		scale only axis,
		xmin=10,
		xmax=50,
		xlabel style={font=\color{white!15!black}, font=\scriptsize},
		xlabel={Number of sensors $M$},
%		xlabel shift = -2.5pt,
		% xmode = log,
%		xtick = {0.025,0.05,0.1,0.2,0.4},
%		xticklabels = {0.025,0.05,0.1,0.2,0.4},
		%ymin=0,
%		ymax=1,
		ylabel style={font=\color{white!15!black}, font=\scriptsize},
		ylabel={Computation time (s)},
%		ylabel shift = -3pt,
		ymode = log,
    ytick distance=10^2,
		axis background/.style={fill=white},
		xmajorgrids,
		ymajorgrids,
		legend style={legend cell align=left, align=left, draw=none,
			% draw=white!15!black, 
			font=\scriptsize, row sep=-2pt,
			legend pos=outer north east,
%			at={(0,0)}, anchor = south west,
		},
		cycle multi list = {
%			solid,dashed,densely dotted\nextlist
%			[2 of]MATLABcolormarklist
			MATLABcolormarklist
%			MATLABcolorlist
		},
		% legend to name=legend2,
		% legend columns=5,
		]

		\addplot table [x index=0, y=SCIP-SDP-N] {\filepath.csv};
		\addlegendentry{MISDP~\eqref{prob:MMVL20MISDP} via SCIP-SDP};
		\addplot[color=MATLABcolor6, densely dashdotted, mark=triangle, every mark/.append style = {solid}] table [x index=0, y=SCIP-SDP-M] {\filepath.csv};
		\addlegendentry{MISDP~\eqref{prob:MMVL20MISDP2} via SCIP-SDP};
		
		\pgfplotsset{cycle list shift=-1};
		
		\addplot table [x index=0, y=Pilanci-N] {\filepath.csv};
		\addlegendentry{MISDP~\eqref{prob:MMVL20MISDP} via RR};
		\addplot[color=MATLABcolor7, densely dashdotted, mark=triangle, every mark/.append style = {solid}] table [x index=0, y=Pilanci-M] {\filepath.csv};
		\addlegendentry{MISDP~\eqref{prob:MMVL20MISDP2} via RR};
		
		\pgfplotsset{cycle list shift=-2};
		
		\addplot table [x index=0, y=SPARROW] {\filepath.csv};
		\addlegendentry{SPARROW};
		
		\addplot table [x index=0, y=SBL] {\filepath.csv};
		\addlegendentry{SBL};
%		\pgfplotsset{cycle list shift=0};
		
		\addplot table [x index=0, y=MUSIC] {\filepath.csv};
		\addlegendentry{MUSIC};

		\addplot+[color=TUDa-4a, densely dashdotted] table [x index=0, y=root-MUSIC] {\filepath.csv};
		\addlegendentry{root-MUSIC};
		
		\addplot[color=TUDa-0c, densely dashdotted] table [x index=0, y=DML] {\filepath.csv};
		\addlegendentry{DML};
		
%		\addplot+[dashed] table [x index=0, y=fmincon] {\filepath};
%		\addlegendentry{DML (\texttt{fmincon})};
    \legend{};
		
	\end{axis}
\end{tikzpicture}
	\caption{Computation time vs. array size for $L=5$ uncorrelated sources, $ N = 50 $ snapshots, $\text{SNR} = -5$ dB, and $K=100$ grid points.}
	\label{fig:varyM_L=5_runtime}
\end{figure}

In \textit{Experiment 5}, we further investigate the computation time of the methods with respect to the array size $M$. We keep the scenario in Experiment 4 with $ L=5 $ sources and fix the number of snapshots $N=50$ while varying the number of sensors $ M $. Fig.~\ref{fig:varyM_L=5_runtime} displays the computation time of different methods averaged over $ N_R = 200 $ Monte-Carlo trials.
The computation time of MISDP~\eqref{prob:MMVL20MISDP2} via SCIP-SDP or the RR algorithm grows more significantly with $ M $ than that of MISDP~\eqref{prob:MMVL20MISDP} since the dimension of the semidefinite constraint in~\eqref{prob:MMVL20MISDP} depends not only on the array size $ M $ but also on the number of snapshots $ N $ that is fixed to $ 50 $.

\begin{figure}[t]
	\centering
	\ref{legend3}

	\subfigure[\label{fig:varyL_noGridless}]{
		\begin{tikzpicture}[inner ysep=1.5pt]
  \def\filepath{results/varyL_M=8_N=8_SNR=10_uncorr_rmse.csv}
	
	\begin{axis}[%
		at={(0,0)},
		scale only axis,
		enlarge x limits={false},
		xlabel style={font=\color{white!15!black}, font=\scriptsize},
    xlabel={Number of sources $ L $},
%		xlabel shift = -2.5pt,
%		xmode = log,
%		xtick = {0.025,0.05,0.1,0.2,0.4},
%		xticklabels = {0.025,0.05,0.1,0.2,0.4},
    % scaled x ticks = {real:3.1415926},
    % xtick scale label code/.code={$\cdot \pi$},
		ymin=1e-2,
		ylabel style={font=\color{white!15!black}, font=\scriptsize},
    ylabel={RMSE (rad)},
%		ylabel shift = -3pt,
		ymode = log,
		axis background/.style={fill=white},
		xmajorgrids,
		ymajorgrids,
		legend style={legend cell align=left, align=left, draw=white, % draw=white!15!black, 
			font=\scriptsize, row sep=-2pt,
			at={(0,0)}, anchor = south west,
      % legend pos=outer north east
    },
		cycle multi list = {
%			solid,dashed,densely dotted\nextlist
%			[2 of]MATLABcolormarklist
			MATLABcolormarklist
%			MATLABcolorlist
		},
    legend to name=legend3,
    legend columns=3,
		]

    \addplot table [x index=0, y={Pilanci(NR=1e4)}] {\filepath};
    \addlegendentry{RR ($ T=10^4 $)};

    \addplot table [x index=0, y={Pilanci(NR=1e5)}] {\filepath};
    \addlegendentry{RR ($ T=10^5 $)};

		\addplot table [x index=0, y=SPARROW] {\filepath};
		\addlegendentry{SPARROW};

		\addplot table [x index=0, y=SBL] {\filepath};
		\addlegendentry{SBL};

		\addplot table [x index=0, y=MUSIC] {\filepath};
		\addlegendentry{MUSIC};

		\addplot+[color=TUDa-4a, densely dashdotted] table [x index=0, y=root-MUSIC] {\filepath};
		\addlegendentry{root-MUSIC};
		
		\addplot[color=black] table [x index=0, y=CRB] {\filepath};
		\addlegendentry{CRB};
		
%		\addplot+[dashed] table [x index=0, y=fmincon] {\filepath};
%		\addlegendentry{DML (\texttt{fmincon})};
		% \legend{};
		
	\end{axis}
\end{tikzpicture}
	}
	\\
	\subfigure[\label{fig:varyL_gridlessDML}]{\begin{tikzpicture}[inner ysep=1.5pt]
  \def\filepath{results/varyL_M=8_N=8_SNR=10_uncorr_fminconDML_rmse.csv}
	
	\begin{axis}[%
		at={(0,0)},
		scale only axis,
		enlarge x limits={false},
		xlabel style={font=\color{white!15!black}, font=\scriptsize},
    xlabel={Number of sources $ L $},
%		xlabel shift = -2.5pt,
%		xmode = log,
%		xtick = {0.025,0.05,0.1,0.2,0.4},
%		xticklabels = {0.025,0.05,0.1,0.2,0.4},
    % scaled x ticks = {real:3.1415926},
    % xtick scale label code/.code={$\cdot \pi$},
		ymin=1e-2,
		ylabel style={font=\color{white!15!black}, font=\scriptsize},
    ylabel={RMSE (rad)},
%		ylabel shift = -3pt,
		ymode = log,
		axis background/.style={fill=white},
		xmajorgrids,
		ymajorgrids,
		legend style={legend cell align=left, align=left, draw=white, % draw=white!15!black, 
			font=\tiny, row sep=-2pt,
			at={(0,0)}, anchor = south west,
      legend pos=outer north east
    },
		cycle multi list = {
%			solid,dashed,densely dotted\nextlist
%			[2 of]MATLABcolormarklist
			MATLABcolormarklist
%			MATLABcolorlist
		},
    % legend to name=legend3,
    % legend columns=8,
		]

    \addplot table [x index=0, y={Pilanci(NR=1e4)}] {\filepath};
    \addlegendentry{RR ($ T=10^4 $)};

    \addplot table [x index=0, y={Pilanci(NR=1e5)}] {\filepath};
    \addlegendentry{RR ($ T=10^5 $)};

		\addplot table [x index=0, y=SPARROW] {\filepath};
		\addlegendentry{SPARROW};

		\addplot table [x index=0, y=SBL] {\filepath};
		\addlegendentry{SBL};

		\addplot table [x index=0, y=MUSIC] {\filepath};
		\addlegendentry{MUSIC};

		\addplot+[color=TUDa-4a, densely dashdotted] table [x index=0, y=root-MUSIC] {\filepath};
		\addlegendentry{root-MUSIC};
		
		\addplot[color=black] table [x index=0, y=CRB] {\filepath};
		\addlegendentry{CRB};
		
%		\addplot+[dashed] table [x index=0, y=fmincon] {\filepath};
%		\addlegendentry{DML (\texttt{fmincon})};
		\legend{};
		
	\end{axis}
\end{tikzpicture}} \\
	\subfigure[\label{fig:varyL_gridlessMAP}]{\begin{tikzpicture}[inner ysep=1.5pt]
  \def\filepath{results/varyL_M=8_N=8_SNR=10_uncorr_fminconMAP_rmse.csv}
	
	\begin{axis}[%
		at={(0,0)},
		scale only axis,
		enlarge x limits={false},
		xlabel style={font=\color{white!15!black}, font=\scriptsize},
    xlabel={Number of sources $ L $},
%		xlabel shift = -2.5pt,
%		xmode = log,
%		xtick = {0.025,0.05,0.1,0.2,0.4},
%		xticklabels = {0.025,0.05,0.1,0.2,0.4},
    % scaled x ticks = {real:3.1415926},
    % xtick scale label code/.code={$\cdot \pi$},
		ymin=1e-2,
		ylabel style={font=\color{white!15!black}, font=\scriptsize},
    ylabel={RMSE (rad)},
%		ylabel shift = -3pt,
		ymode = log,
		axis background/.style={fill=white},
		xmajorgrids,
		ymajorgrids,
		legend style={legend cell align=left, align=left, draw=white, % draw=white!15!black, 
			font=\tiny, row sep=-2pt,
			at={(0,0)}, anchor = south west,
      legend pos=outer north east
    },
		cycle multi list = {
%			solid,dashed,densely dotted\nextlist
%			[2 of]MATLABcolormarklist
			MATLABcolormarklist
%			MATLABcolorlist
		},
    % legend to name=legend3,
    % legend columns=8,
		]

    \addplot table [x index=0, y={Pilanci(NR=1e4)}] {\filepath};
    \addlegendentry{RR ($ T=10^4 $)};

    \addplot table [x index=0, y={Pilanci(NR=1e5)}] {\filepath};
    \addlegendentry{RR ($ T=10^5 $)};

		\addplot table [x index=0, y=SPARROW] {\filepath};
		\addlegendentry{SPARROW};

		\addplot table [x index=0, y=SBL] {\filepath};
		\addlegendentry{SBL};

		\addplot table [x index=0, y=MUSIC] {\filepath};
		\addlegendentry{MUSIC};

		\addplot+[color=TUDa-4a, densely dashdotted] table [x index=0, y=root-MUSIC] {\filepath};
		\addlegendentry{root-MUSIC};
		
		\addplot[color=black] table [x index=0, y=CRB] {\filepath};
		\addlegendentry{CRB};
		
%		\addplot+[dashed] table [x index=0, y=fmincon] {\filepath};
%		\addlegendentry{DML (\texttt{fmincon})};
		\legend{};
		
	\end{axis}
\end{tikzpicture}}
	\caption{RMSE vs. the number of sources $ L $ for uncorrelated sources, $M=8$ sensors, $N=8$ snapshots, SNR $ = 10 $ dB, and $K=1000$ grid points, (a) without gridless local search, (b) with gridless local search on the DML function, (c) with gridless local search on the MAP function.}
	\label{fig:varyL_rmse}
\end{figure}

\begin{figure}[t]
	\centering
	\ref{legend3_a}

	\begin{tikzpicture}[inner ysep=1.5pt]
  \def\filepath{results/varyL_M=8_N=8_SNR=10_uncorr_runtime.csv}
	
	\begin{axis}[%
		at={(0,0)},
		scale only axis,
		enlarge x limits={false},
		xlabel style={font=\color{white!15!black}, font=\scriptsize},
    xlabel={Number of sources $ L $},
%		xlabel shift = -2.5pt,
%		xmode = log,
%		xtick = {0.025,0.05,0.1,0.2,0.4},
%		xticklabels = {0.025,0.05,0.1,0.2,0.4},
    % scaled x ticks = {real:3.1415926},
    % xtick scale label code/.code={$\cdot \pi$},
		%ymin=0,
%		ymax=1,
		ylabel style={font=\color{white!15!black}, font=\scriptsize},
    ylabel={Computation time (s)},
%		ylabel shift = -3pt,
		ymode = log,
		axis background/.style={fill=white},
		xmajorgrids,
		ymajorgrids,
		legend style={legend cell align=left, align=left, draw=white, % draw=white!15!black, 
			font=\scriptsize, row sep=-2pt,
			at={(0,0)}, anchor = south west,
      legend pos=outer north east
    },
		cycle multi list = {
%			solid,dashed,densely dotted\nextlist
%			[2 of]MATLABcolormarklist
			MATLABcolormarklist
%			MATLABcolorlist
		},
    legend to name=legend3_a,
    legend columns=3,
		]

    \addplot table [x index=0, y={Pilanci(NR=1e4)}] {\filepath};
    \addlegendentry{RR ($ T=10^4 $)};

    \addplot table [x index=0, y={Pilanci(NR=1e5)}] {\filepath};
    \addlegendentry{RR ($ T=10^5 $)};

		\addplot table [x index=0, y=SPARROW] {\filepath};
		\addlegendentry{SPARROW};

		\addplot table [x index=0, y=SBL] {\filepath};
		\addlegendentry{SBL};

		\addplot table [x index=0, y=MUSIC] {\filepath};
		\addlegendentry{MUSIC};

		\addplot+[color=TUDa-4a, densely dashdotted] table [x index=0, y=root-MUSIC] {\filepath};
		\addlegendentry{root-MUSIC};
		
		% \addplot[color=black] table [x index=0, y=CRB] {\filepath};
		% \addlegendentry{CRB};
		
%		\addplot+[dashed] table [x index=0, y=fmincon] {\filepath};
%		\addlegendentry{DML (\texttt{fmincon})};
		% \legend{};
		
	\end{axis}
\end{tikzpicture}
	\caption{Computation time vs. the number of sources $ L $ for uncorrelated sources, $M=8$ sensors, $N=8$ snapshots, SNR $ = 10 $ dB, and $K=1000$ grid points.}
	\label{fig:varyL_runtime}
\end{figure}

In \textit{Experiment 6}, we fix SNR $=10$ dB and $ N = 8 $ and vary $ L $ from $1$ to $M-1$. The $ L $ sources are uniformly spaced with frequencies $\vect{\mu}_l = \pi \cdot \left( (2 (l-1) + 1) / L - 0.895 \right) $ for $ l = 1, \ldots, L $.
The brute-force DML and SCIP-SDP are excluded in this experiment, as well as in Experiment 7, due to their high computation time for large $L$ or $ K $.
Thus, we can further increase the grid size to $ K = 1000 $ to reduce the effect of grid mismatch error in the inspected region of $ L $.
The estimation error and computation time are reported in Fig.~\ref{fig:varyL_rmse} and~\ref{fig:varyL_runtime}, respectively.
We examine both $ T = 10^4 $ and $ T = 10^5 $ for the RR algorithm since a suitable choice of $ T $ for balancing the estimation quality and the computational complexity depends on the size of the binary feasible set, which increases exponentially with $ L $. We remark that further increasing $T$ does no longer provide a significant improvement on the estimation quality for the considered problem size.

In Fig.~\ref{fig:varyL_rmse}, MISDP exhibits superior error performance for large $ L $ compared to the other methods, especially for $L=M-1$, where all other methods fail.
Moreover, when an additional local search is performed, the superior error performance of the RR algorithm is achieved with a small $ T=10^4 $, which shows a computation time comparable to that of SPARROW.
This demonstrates the significantly improved scalability of the randomized rounding strategy compared to the exact branch-and-bound approach, whose complexity grows exponentially with problem dimensions.
The computation time of SPARROW grows with $ L $ because of the coordinate descent implementation, whereas the computation time of the other methods remains almost constant.

\begin{figure}[t]
	\centering
	\ref{legend_varyK_runtime}

	\begin{tikzpicture}[inner ysep=1.5pt]
  \def\filepath{results/varyK_L=5_SNR=-5_N=100_uncorr_runtime.csv}
	
	\begin{axis}[%
		at={(0,0)},
		scale only axis,
		enlarge x limits={false},
		xlabel style={font=\color{white!15!black}, font=\scriptsize},
    xlabel={Grid size $ K $},
%		xlabel shift = -2.5pt,
    xmode = log,
%		xtick = {0.025,0.05,0.1,0.2,0.4},
%		xticklabels = {0.025,0.05,0.1,0.2,0.4},
    % scaled x ticks = {real:3.1415926},
    % xtick scale label code/.code={$\cdot \pi$},
		%ymin=0,
%		ymax=1,
		ylabel style={font=\color{white!15!black}, font=\scriptsize},
    ylabel={Computation time (s)},
%		ylabel shift = -3pt,
		ymode = log,
		axis background/.style={fill=white},
		xmajorgrids,
		ymajorgrids,
		legend style={legend cell align=left, align=left, draw=white, % draw=white!15!black, 
			font=\scriptsize, row sep=-2pt,
			at={(0,0)}, anchor = south west,
      % legend pos=outer north,
    },
		cycle multi list = {
%			solid,dashed,densely dotted\nextlist
%			[2 of]MATLABcolormarklist
			MATLABcolormarklist
%			MATLABcolorlist
		},
    legend to name=legend_varyK_runtime,
    legend columns=3,
		]

    \addplot table [x index=0, y=Pilanci-M] {\filepath};
    \addlegendentry{MISDP~\eqref{prob:MMVL20MISDP2} via RR};

		\addplot[color=MATLABcolor6, densely dashdotted, mark=triangle, every mark/.append style = {solid}] table [x index=0, y=IR] {\filepath};
		\addlegendentry{interval relaxation};
		
		\pgfplotsset{cycle list shift=-1};

    \addplot table [x index=0, y=RR] {\filepath};
    \addlegendentry{randomized rounding};

		\addplot table [x index=0, y=SPARROW] {\filepath};
		\addlegendentry{SPARROW};

		\addplot table [x index=0, y=SBL] {\filepath};
		\addlegendentry{SBL};

		\addplot table [x index=0, y=MUSIC] {\filepath};
		\addlegendentry{MUSIC};

		\addplot+[color=TUDa-4a, densely dashdotted] table [x index=0, y=root-MUSIC] {\filepath};
		\addlegendentry{root-MUSIC};
		
		% \addplot[color=black] table [x index=0, y=CRB] {\filepath};
		% \addlegendentry{CRB};
		
%		\addplot+[dashed] table [x index=0, y=fmincon] {\filepath};
%		\addlegendentry{DML (\texttt{fmincon})};
		% \legend{};
		
	\end{axis}
\end{tikzpicture}
	\caption{Computation time vs. the grid size $ K $ for $ L=5 $ uncorrelated sources, $M=8$ sensors, $N=100$ snapshots, SNR $ = 10 $ dB.}
	\label{fig:varyK_runtime}
\end{figure}

In \textit{Experiment 7}, we evaluate the scalability of different methods with respect to the grid size $ K $. We consider the scenario in Experiment 4 with $ L=5 $ uncorrelated sources and fix SNR $ =10 $ dB and $N=100$ while varying the grid size $ K $. The computation time is reported in Fig.~\ref{fig:varyK_runtime} averaged over $ N_R = 200 $ Monte-Carlo trials. The computation time of the two steps of the RR algorithm, namely, solving the interval relaxation and performing randomized rounding, are also displayed separately.
To exclude the influence of $N$ on the computation time of MISDP in this experiment, we only consider the formulation~\eqref{prob:MMVL20MISDP2}.

In Fig.~\ref{fig:varyK_runtime}, the computation time of the RR algorithm is dominated by the randomized rounding step for small $ K $, whereas the complexity of the interval relaxation grows greatly with $ K $ and becomes the major part of the total computation time for large $ K $.
A customized algorithm for solving the SDP interval relaxation that is more scalable than the interior-point solvers may significantly reduce the computation time of the RR algorithm for large $ K $. This may potentially be achieved by primal-dual methods such as ADMM.
The scalability of the randomized rounding step is comparable to that of SBL, both of which are superior to the scalability of SPARROW.

\subsection{Correlated Source Signals}
\label{subsec:correlated}
In \textit{Experiment 8}, we evaluate the performance of the methods in the case with correlated source signals. We consider $L=3$ sources with frequencies $\vect{\mu} = \pi \cdot [-0.1, 0.35, 0.47]^\Trans$. The source signals follow a zero-mean complex Gaussian distribution with the covariance matrix
\begin{equation}
	\label{eq:covmat}
	\begin{bmatrix}
		1 & \varphi & \varphi \\
		\varphi^* & 1 & \varphi^2 \\
		\varphi^* & \varphi^{*2} & 1
	\end{bmatrix}
\end{equation}
and the correlation coefficient $\varphi \in \Compl$ between source 1 and 2 is chosen to be $\varphi=0.99$. Since all sources have the same average power of $1$, the regularization $\rho$ is chosen according to the rule~\eqref{eq:rho} with $P_\Psi = 1$. The estimation errors for various choices of the number of snapshots are displayed in Fig.~\ref{fig:varyN_L=3_corr}.

The following differences can be observed compared to the results in Fig.~\ref{fig:varyN_L=3} for uncorrelated sources.
As the sources are well separated, the CRB is not very sensitive to the correlation. Nevertheless, the high correlation causes a notable performance degradation of all methods.
% It is verified that the subspace-based methods, including MUSIC and root-MUSIC, often fail for correlated sources. The SBL method exhibits a significant degradation of the error performance due to the mismatch of the prior model. Although the RMSE of the brute-force DML and the proposed method also increases in the post-threshold region, the increase in RMSE is insignificant.
% and the proposed MISDP method demonstrates its superior robustness to the source correlation compared to other methods
Apart from DML and the proposed MISDP method, all the other methods fail to recover the three DOAs in the inspected region of $N$.
Although the MAP estimation model in the MISDP method incorporates an uncorrelated prior assumption, MISDP shows superior robustness to the source correlation in Fig.~\ref{fig:varyN_L=3_corr} as the increase of its RMSE compared to that in Fig.~\ref{fig:varyN_L=3} is marginal.

%\begin{figure*}[t]
%	\centering
%	\ref{legend1}
%	
%	\subfigure[no gridless local search]{
%		\input{images/varySNR_L=3_N=8_corr_rmse.tikz}
%	}
%	\subfigure[gridless local search on DML]{\input{images/varySNR_L=3_N=8_corr_fminconDML_rmse.tikz}}
%	\subfigure[gridless local search on MAP]{\input{images/varySNR_L=3_N=8_corr_fminconMAP_rmse.tikz}}
%	\caption{RMSE vs. SNR for $L=3$ correlated sources, $M=8$ sensors, $N=8$ snapshots, and $K=100$ grid points.}
%	%	\label{fig:varySNR_L=3}
%\end{figure*}

\begin{figure}[t]
	\centering
	\ref{legend1}

	\subfigure[]{\begin{tikzpicture}[inner ysep=1.5pt]
	\def\filepath{results/varyN_L=3_SNR=-5_corr_rmse.csv}
	
	\begin{axis}[%
		at={(0,0)},
		scale only axis,
		enlarge x limits={false},
		xlabel style={font=\color{white!15!black}, font=\scriptsize},
		xlabel={Number of snapshots $N$},
%		xlabel shift = -2.5pt,
		xmode = log,
		ylabel style={font=\color{white!15!black}, font=\scriptsize},
		ylabel={RMSE (rad)},
%		ylabel shift = -3pt,
		ymode = log,
		axis background/.style={fill=white},
		xmajorgrids,
		ymajorgrids,
		legend style={legend cell align=left, align=left, draw=white, % draw=white!15!black, 
			font=\tiny, row sep=-2pt,
			at={(0,0)}, anchor = south west},
		cycle multi list = {
%			solid,dashed,densely dotted\nextlist
%			[2 of]MATLABcolormarklist
			MATLABcolormarklist
%			MATLABcolorlist
		},
%		legend to name=legend1,
%		legend columns=8,
		]

		\addplot table [x index=0, y=SCIP-SDP] {\filepath};
		\addlegendentry{MISDP via SCIP-SDP};
		
%		\addplot[color=MATLABcolor6, densely dashdotted, mark=triangle, every mark/.append style = {solid}] table [x index=0, y=SCIP-SDP-SVD] {\filepath};
%		\addlegendentry{MISDP-S via SCIP-SDP};
%		
%		\pgfplotsset{cycle list shift=-1};
		
		\addplot table [x index=0, y=Pilanci] {\filepath};
		\addlegendentry{MISDP via IRRR};
		
%		\addplot[color=MATLABcolor7, densely dashdotted, mark=triangle, every mark/.append style = {solid}] table [x index=0, y=Pilanci-SVD] {\filepath};
%		\addlegendentry{MISDP-S via IRRR};
%		
%		\pgfplotsset{cycle list shift=-2};
		
		\addplot table [x index=0, y=SPARROW] {\filepath};
		\addlegendentry{SPARROW};
		
		\addplot table [x index=0, y=SBL] {\filepath};
		\addlegendentry{SBL};
		
		\addplot table [x index=0, y=MUSIC] {\filepath};
		\addlegendentry{MUSIC};
		
		\addplot+[color=TUDa-4a, densely dashdotted] table [x index=0, y=root-MUSIC] {\filepath};
		\addlegendentry{root-MUSIC};
		
		\addplot+[color=TUDa-0c, densely dashdotted,mark=null] table [x index=0, y=DML] {\filepath};
		\addlegendentry{DML};
		
		\addplot[color=black] table [x index=0, y=CRB] {\filepath};
		\addlegendentry{CRB};
		
%		\addplot+[dashed] table [x index=0, y=fmincon] {\filepath};
%		\addlegendentry{DML (\texttt{fmincon})};
		\legend{};
		
	\end{axis}
\end{tikzpicture}} \\
	\subfigure[]{\begin{tikzpicture}[inner ysep=1.5pt]
	\def\filepath{results/varyN_L=3_SNR=-5_corr_fminconDML_rmse.csv}
	
	\begin{axis}[%
		at={(0,0)},
		scale only axis,
		enlarge x limits={false},
		xlabel style={font=\color{white!15!black}, font=\scriptsize},
		xlabel={Number of snapshots $N$},
%		xlabel shift = -2.5pt,
		xmode = log,
		ylabel style={font=\color{white!15!black}, font=\scriptsize},
		ylabel={RMSE (rad)},
%		ylabel shift = -3pt,
		ymode = log,
		axis background/.style={fill=white},
		xmajorgrids,
		ymajorgrids,
		legend style={legend cell align=left, align=left, draw=white, % draw=white!15!black, 
			font=\tiny, row sep=-2pt,
			at={(0,0)}, anchor = south west},
		cycle multi list = {
%			solid,dashed,densely dotted\nextlist
%			[2 of]MATLABcolormarklist
			MATLABcolormarklist
%			MATLABcolorlist
		},
%		legend to name=legend1,
%		legend columns=8,
		]

		\addplot table [x index=0, y=SCIP-SDP] {\filepath};
		\addlegendentry{MISDP via SCIP-SDP};
		
%		\addplot[color=MATLABcolor6, densely dashdotted, mark=triangle, every mark/.append style = {solid}] table [x index=0, y=SCIP-SDP-SVD] {\filepath};
%		\addlegendentry{MISDP-S via SCIP-SDP};
%		
%		\pgfplotsset{cycle list shift=-1};
		
		\addplot table [x index=0, y=Pilanci] {\filepath};
		\addlegendentry{MISDP via IRRR};
		
%		\addplot[color=MATLABcolor7, densely dashdotted, mark=triangle, every mark/.append style = {solid}] table [x index=0, y=Pilanci-SVD] {\filepath};
%		\addlegendentry{MISDP-S via IRRR};
%		
%		\pgfplotsset{cycle list shift=-2};
		
		\addplot table [x index=0, y=SPARROW] {\filepath};
		\addlegendentry{SPARROW};
		
		\addplot table [x index=0, y=SBL] {\filepath};
		\addlegendentry{SBL};
		
		\addplot table [x index=0, y=MUSIC] {\filepath};
		\addlegendentry{MUSIC};
		
		\addplot+[color=TUDa-4a, densely dashdotted] table [x index=0, y=root-MUSIC] {\filepath};
		\addlegendentry{root-MUSIC};
		
		\addplot+[color=TUDa-0c, densely dashdotted,mark=null] table [x index=0, y=DML] {\filepath};
		\addlegendentry{DML};
		
		\addplot[color=black] table [x index=0, y=CRB] {\filepath};
		\addlegendentry{CRB};
		
%		\addplot+[dashed] table [x index=0, y=fmincon] {\filepath};
%		\addlegendentry{DML (\texttt{fmincon})};
		\legend{};
		
	\end{axis}
\end{tikzpicture}} \\
	\subfigure[]{\begin{tikzpicture}[inner ysep=1.5pt]
	\def\filepath{results/varyN_L=3_SNR=-5_corr_fminconMAP_rmse.csv}
	
	\begin{axis}[%
		at={(0,0)},
		scale only axis,
		enlarge x limits={false},
		xlabel style={font=\color{white!15!black}, font=\scriptsize},
		xlabel={Nnumber of snapshots $N$},
%		xlabel shift = -2.5pt,
		xmode = log,
		ylabel style={font=\color{white!15!black}, font=\scriptsize},
		ylabel={RMSE (rad)},
%		ylabel shift = -3pt,
		ymode = log,
		axis background/.style={fill=white},
		xmajorgrids,
		ymajorgrids,
		legend style={legend cell align=left, align=left, draw=white, % draw=white!15!black, 
			font=\tiny, row sep=-2pt,
			at={(0,0)}, anchor = south west},
		cycle multi list = {
%			solid,dashed,densely dotted\nextlist
%			[2 of]MATLABcolormarklist
			MATLABcolormarklist
%			MATLABcolorlist
		},
%		legend to name=legend1,
%		legend columns=8,
		]

		\addplot table [x index=0, y=SCIP-SDP] {\filepath};
		\addlegendentry{MISDP via SCIP-SDP};
		
%		\addplot[color=MATLABcolor6, densely dashdotted, mark=triangle, every mark/.append style = {solid}] table [x index=0, y=SCIP-SDP-SVD] {\filepath};
%		\addlegendentry{MISDP-S via SCIP-SDP};
%		
%		\pgfplotsset{cycle list shift=-1};
		
		\addplot table [x index=0, y=Pilanci] {\filepath};
		\addlegendentry{MISDP via IRRR};
		
%		\addplot[color=MATLABcolor7, densely dashdotted, mark=triangle, every mark/.append style = {solid}] table [x index=0, y=Pilanci-SVD] {\filepath};
%		\addlegendentry{MISDP-S via IRRR};
%		
%		\pgfplotsset{cycle list shift=-2};
		
		\addplot table [x index=0, y=SPARROW] {\filepath};
		\addlegendentry{SPARROW};
		
		\addplot table [x index=0, y=SBL] {\filepath};
		\addlegendentry{SBL};
		
		\addplot table [x index=0, y=MUSIC] {\filepath};
		\addlegendentry{MUSIC};
		
		\addplot+[color=TUDa-4a, densely dashdotted] table [x index=0, y=root-MUSIC] {\filepath};
		\addlegendentry{root-MUSIC};
		
		\addplot+[color=TUDa-0c, densely dashdotted,mark=null] table [x index=0, y=DML] {\filepath};
		\addlegendentry{DML};
		
		\addplot[color=black] table [x index=0, y=CRB] {\filepath};
		\addlegendentry{CRB};
		
%		\addplot+[dashed] table [x index=0, y=fmincon] {\filepath};
%		\addlegendentry{DML (\texttt{fmincon})};
		\legend{};
		
	\end{axis}
\end{tikzpicture}}
	\caption{Error performance w.r.t. the number of snapshots for $L=3$ correlated sources, $M=8$ sensors, $\text{SNR} = -5$ dB, and $K=100$ grid points, in the case (a) without gridless local search, (b) with gridless local search on the DML function, (c) with gridless local search on the MAP function.}
	\label{fig:varyN_L=3_corr}
\end{figure}

\subsection{Sensitivity to Regularization Parameter}

\begin{figure}[t]
	\begin{center}
		\ref{legend4}

		\subfigure[]{\begin{tikzpicture}[inner ysep=1.5pt]
	\def\filepath{results}
	
	\begin{axis}[%
		% width=0.35\columnwidth,
		%height=2cm,
		at={(0,0)},
		scale only axis,
		enlarge x limits={false},
		xlabel style={font=\color{white!15!black}, font=\scriptsize},
		xlabel={$ \rho / (\sigma^2/\gamma) $},
%		xlabel shift = -2.5pt,
		xmode = log,
		ylabel style={font=\color{white!15!black}, font=\scriptsize},
		ylabel={RMSE (rad)},
%		ylabel shift = -3pt,
		ymode = log,
		axis background/.style={fill=white},
		xmajorgrids,
		ymajorgrids,
		legend style={legend cell align=left, align=left, draw=none, % draw=white!15!black, 
			font=\scriptsize, row sep=-2pt,
			% at={(0.05,0.2)}, anchor = south west,
    },
		cycle multi list = {
%			solid,dashed,densely dotted\nextlist
%			[2 of]MATLABcolormarklist
			MATLABcolormarklist
%			MATLABcolorlist
		},
		legend to name=legend4,
		legend columns=2,
		]

		% \addplot table [x index=0, y=SCIP-SDP] {\filepath};
		% \addlegendentry{MISDP via SCIP-SDP};
		
%		\addplot[color=MATLABcolor6, densely dashdotted, mark=triangle, every mark/.append style = {solid}] table [x index=0, y=SCIP-SDP-SVD] {\filepath};
%		\addlegendentry{MISDP-S via SCIP-SDP};
%		
%		\pgfplotsset{cycle list shift=-1};

    \addplot table [x index=0, y=Pilanci] {\filepath/varyPenalty_L=5_SNR=-5_N=100_uncorr_rmse.csv};
    \addlegendentry{$N=100$, SNR $= -5$ dB};
		\addplot table [x index=0, y=Pilanci] {\filepath/varyPenalty_L=5_SNR=-5_N=8_uncorr_rmse.csv};
    \addlegendentry{$N=8$, SNR $= -5$ dB};
		\addplot table [x index=0, y=Pilanci] {\filepath/varyPenalty_L=5_SNR=5_N=8_uncorr_rmse.csv};
    \addlegendentry{$N=8$, SNR $= 5$ dB};
		
%		\addplot[color=MATLABcolor7, densely dashdotted, mark=triangle, every mark/.append style = {solid}] table [x index=0, y=Pilanci-SVD] {\filepath};
%		\addlegendentry{MISDP-S via IRRR};
%		
%		\pgfplotsset{cycle list shift=-2};
		
		% \addplot table [x index=0, y=SPARROW] {\filepath};
		% \addlegendentry{SPARROW};
		% 
		% \addplot table [x index=0, y=SBL] {\filepath};
		% \addlegendentry{SBL};
		% 
		% \addplot table [x index=0, y=MUSIC] {\filepath};
		% \addlegendentry{MUSIC};
		%
		% \addplot+[color=TUDa-4a, densely dashdotted] table [x index=0, y=root-MUSIC] {\filepath};
		% \addlegendentry{root-MUSIC};
		% 
		% \addplot+[color=TUDa-0c, densely dashdotted,mark=null] table [x index=0, y=DML] {\filepath};
		% \addlegendentry{DML};

  %   \addplot[color=black,dashed] table [x index=0, y=CRB] {\filepath/varyPenalty_L=3_SNR=5_N=8_uncorr_rmse.csv};
		% \addlegendentry{CRB};
		
%		\addplot+[dashed] table [x index=0, y=fmincon] {\filepath};
%		\addlegendentry{DML (\texttt{fmincon})};
	% \legend{};
		
	\end{axis}
\end{tikzpicture}}
		\\
		\subfigure[]{\begin{tikzpicture}[inner ysep=1.5pt]
	\def\filepath{results}
	
	\begin{axis}[%
    % width=0.35\columnwidth,
		%height=2cm,
		at={(0,0)},
		scale only axis,
    enlarge x limits={false},
		xlabel style={font=\color{white!15!black}, font=\scriptsize},
		xlabel={$ \rho / (\sigma^2/\gamma) $},
%		xlabel shift = -2.5pt,
		xmode = log,
		ylabel style={font=\color{white!15!black}, font=\scriptsize},
		ylabel={Computation time (s)},
%		ylabel shift = -3pt,
		ymode = log,
		axis background/.style={fill=white},
		xmajorgrids,
		ymajorgrids,
		legend style={legend cell align=left, align=left, draw,
			% draw=white!15!black, 
			font=\scriptsize, row sep=-2pt,
			legend pos=outer north east,
%			at={(0,0)}, anchor = south west,
		},
		cycle multi list = {
%			solid,dashed,densely dotted\nextlist
%			[2 of]MATLABcolormarklist
			MATLABcolormarklist
%			MATLABcolorlist
		},
		% legend to name=legend2,
		% legend columns=2,
		]

		\addplot table [x index=0, y=Pilanci] {\filepath/varyPenalty_L=5_SNR=-5_N=100_uncorr_runtime.csv};
    \addlegendentry{$N=100$, SNR $= -5$ dB};	
		\addplot table [x index=0, y=Pilanci] {\filepath/varyPenalty_L=5_SNR=-5_N=8_uncorr_runtime.csv};
    \addlegendentry{$N=8$, SNR $= -5$ dB};
    \addplot table [x index=0, y=Pilanci] {\filepath/varyPenalty_L=5_SNR=5_N=8_uncorr_runtime.csv};
    \addlegendentry{$N=8$, SNR $= 5$ dB};
    \legend{};
				
	\end{axis}
\end{tikzpicture}}
	\end{center}
	\caption{Sensitivity of MISDP via RR to the regularization parameter $ \rho $ for $ L=5 $ uncorrelated sources, $ M=8 $ sensors, and $ K = 100 $ grid points.}\label{fig:sensitivity}
\end{figure}

In the previous simulations, the regularization parameter $ \rho $ for MISDP is chosen by the rule~\eqref{eq:rho} derived from the MAP estimation. However, the source and noise variances are often unknown in practice, which makes the proper choice of $ \rho $ difficult. In \textit{Experiment 9}, we investigate the sensitivity of MISDP via the RR algorithm to the choice of $ \rho $.
Both the estimation error and the computation time are reported in Fig.~\ref{fig:sensitivity} with $ \rho $ varied from $ 0.01 \cdot \sigma^2 / \gamma $ to $ 100 \cdot \sigma^2 / \gamma $.
We consider the scenario in Experiment 4 with $ L=5 $ uncorrelated sources. Two values of $ N $ before and after the threshold in Fig.~\ref{fig:varyN_L=5}, respectively, are considered. As shown in Fig.~\ref{fig:sensitivity}, for $ \rho < \sigma^2 / \gamma $, which is the value determined by the MAP estimation, the estimation error of the proposed method varies marginally and its computation time is relatively high. For a small $ \rho $, the computational complexity of both the interval relaxation and the evaluation of the objective function, which involves the inversion of a nearly singular matrix, in the randomized rounding step increases significantly.
For $ \rho > \sigma^2 / \gamma $, the computation time is reduced but the estimation error increases dramatically due to the increase of the bias caused by the regularization. This observation suggests the following heuristic guideline for choosing the regularization parameter $ \rho $. A proper value of $ \rho $ can be found by successively decreasing $ \rho $ until the computation time increases significantly.

\section{Conclusion}
\label{sec:conclusion}

We propose a maximum a posteriori (MAP) estimator for DOA estimation from multiple snapshots, reformulated as a mixed-integer semidefinite program (MISDP).
This enables efficient computation of globally optimal solutions using state-of-the-art branch-and-bound solvers, with the added benefit of optimality assessment even if terminated early.
To further improve scalability, we introduce a randomized rounding algorithm that finds high-quality approximate solutions of the MISDP problem at significantly reduced computation time for large-scale problems.
Numerical simulations demonstrate superior performance of the proposed MISDP-based method in terms of threshold behavior, resolution, and robustness to heavy source correlations compared to several widely used DOA estimation methods.

Although the effectiveness of randomized rounding is shown by the numerical simulations, its performance highly depends on the quality of the solution of a continuous relaxation of MISDP.
A theoretical guarantee of the tightness of the continuous relaxation for the considered array manifold is still lacking and further investigation is needed in future work.

\bibliographystyle{IEEEtran}
\bibliography{refsPilanci}

\end{document}